\documentclass[useAMS,usenatbib]{mn2e}

\usepackage{lscape,graphicx,amssymb,aas_macros}
\usepackage{amsmath} 
\usepackage{amssymb} 
\usepackage{graphicx} 
\usepackage{indentfirst} 
\usepackage{pdflscape} 
\usepackage{placeins} 
\usepackage{nicefrac} 
\usepackage{afterpage} 
\usepackage{booktabs} 
\usepackage{color} 
\usepackage{float}

\newcommand{\apg}{\:^{>}_{\sim}\:}
\newcommand{\apl}{\:^{<}_{\sim}\:}
\newcommand{\cmjj}{\mbox{${\rm cm^{-2}}$}}
\newcommand{\cmjjj}{\mbox{${\rm cm^{-3}}$}}
\newcommand{\etal}{et al.}

\newcommand{\kms}{\mbox{km\ s${^{-1}}$}}
\newcommand{\lya}{\mbox{${\rm Ly}\alpha$}}

\title[ The radial profile of gas-phase Fe/$\alpha$ ratio around distant galaxies]{On the radial profile of gas-phase Fe/$\alpha$ ratio around distant galaxies} 
\author[Zahedy et al.]{Fakhri S. Zahedy$^{1}$\thanks{E-mail:
fsz@uchicago.edu}, Hsiao-Wen Chen$^{1,2}$\thanks{E-mail:
hchen@oddjob.uchicago.edu}, Jean-Ren\'{e} Gauthier$^{3}$, Michael Rauch$^{4}$ \\
$^{1}$Department of Astronomy \& Astrophysics, The University of Chicago, Chicago, IL 60637, USA \\
$^{2}$Kavli Institute for Cosmological Physics, The University of Chicago, Chicago, IL 60637, USA  \\
$^{3}$DataScience, Inc., Culver City, CA, USA \\
$^{4}$The Observatories of the Carnegie Institution for Science, 813 Santa Barbara Street, Pasadena, CA 91101, USA}

\begin{document}

\pagerange{\pageref{firstpage}--\pageref{lastpage}} \pubyear{2002}

\maketitle

\label{firstpage}

\begin{abstract}

This paper presents a study of the chemical compositions in cool gas
around a sample of 27 intermediate-redshift galaxies.  The sample
comprises 13 massive quiescent galaxies at $z=0.40-0.73$ probed by QSO
sightlines at projected distances $d=3-400$ kpc, and 14 star-forming
galaxies at $z=0.10-1.24$ probed by QSO sightlines at $d=8-163$ kpc.
The main goal of this study is to examine the radial profiles of the
gas-phase Fe/$\alpha$ ratio in galaxy halos based on the observed
Fe\,II to Mg\,II  column density ratios. Because $\mathrm{Mg^+}$ and
$\mathrm{Fe^+}$ share similar ionization potentials, the relative ionization correction is small in moderately ionized gas and the observed ionic abundance ratio $\mathrm{\mathit{N}(Fe\,II)/\mathit{N}(Mg\,II)}$ places a lower limit to the underlying (Fe/Mg) elemental abundance ratio. For quiescent galaxies, a median and dispersion of
$\mathrm{log\,\langle \mathit{N}(Fe\,II)/\mathit{N}(Mg\,II)}\rangle_{\rm med}= -0.06\pm 0.15$ is found at $d\apl 60$ kpc, which declines to
$\mathrm{log\,\langle \mathit{N}(Fe\,II)/\mathit{N}(Mg\,II)}\rangle_{\rm med}<-0.3$ at $d\apg100$ kpc.  On the other hand, star-forming galaxies exhibit
$\mathrm{log\,\langle \mathit{N}(Fe\,II)/\mathit{N}(Mg\,II)}\rangle=
-0.25\pm 0.21$ at $d\apl 60$ kpc and $\mathrm{log\,\langle
  \mathit{N}(Fe\,II)/\mathit{N}(Mg\,II)}\rangle= -0.9\pm 0.4$ at
larger distances. Including possible differential dust depletion or ionization correction would only increase the inferred (Fe/Mg) ratio. The observed $\mathrm{ \mathit{N}(Fe\,II)/\mathit{N}(Mg\,II)}$ implies super-solar Fe/$\alpha$ ratios in the inner halo of quiescent galaxies.  An enhanced Fe abundance indicates a substantial contribution by Type Ia supernovae in the chemical enrichment, which is at least comparable to what is observed in the solar neighborhood or in intracluster media but differs from young star-forming regions.  In the outer halos of quiescent galaxies and in halos around star-forming galaxy, however, the observed $\mathrm{\mathit{N}(Fe\,II)/\mathit{N}(Mg\,II)}$ is consistent with an $\alpha$-element enhanced enrichment pattern, suggesting a core-collapse supernovae dominated enrichment history.  
\end{abstract}


\section{Introduction}

The presence of chemically-enriched gas out to large projected
distances $d\sim 100$ kpc from galaxies is commonly attributed to
super-galactic winds in starburst galaxies (e.g., Murray \etal\ 2011;
Booth \etal\ 2013, Borthakur \etal\ 2013).  However, the presence of
chemically-enriched cool gas around quiescent galaxies both in the
local universe and at high redshifts (e.g., Young \etal\ 2011;
Gauthier \& Chen 2011; Zhu \etal\ 2014; Huang \etal\ 2016) is
difficult to reconcile on the basis of a simple outflow model.
Because of a lack of intense star formation for $\apg 1$ Gyrs in these
quiescent galaxies, additional mechanisms beyond super-galactic winds
are clearly needed to explain the presence of chemically enriched gas
around galaxies.  While tidal interactions and ram pressure force can
work to remove interstellar gas of satellite galaxies and fill the
halo around the primary galaxy (e.g., Wang 1993; Agertz \etal\ 2009;
Gauthier 2013), without a continuing feedback mechanism the gas is
expected to cool and trigger new generations of star formation in the
center of the galaxy (e.g., Conroy \etal\ 2015).

In a recent study, Zahedy \etal\ (2016) investigated the cool gas
content around three lensing galaxies at redshift $z\sim 0.5$, using
multiply-lensed QSOs.  These lensing galaxies are massive with total
stellar mass of $\log\,M_*/M_\odot\apg 11$ and half-light radii of
$r_e = 2.6-8$ kpc.  Their spectral and photometric properties resemble
nearby elliptical galaxies, showing no trace of on-going star
formation.  The multiply-lensed images of the background QSOs occur at
$d= 3-15$ kpc, corresponding to $1-2 \,r_e$, and therefore provide a
sensitive probe of both interstellar and circumgalactic gas in these
quiescent, lensing galaxies.  An interesting finding from this study
is that wherever $\mathrm{Mg\,II}$ absorption transitions are
detected, strong $\mathrm{Fe\,II}$ absorption features are also found
with an observed column density ratio,
$\mathrm{\mathit{N}(Fe\,II)/\mathit{N}(Mg\,II)}$, exceeding the
typical values seen along random QSO sightlines (e.g., Rigby
\etal\ 2002; Narayanan \etal\ 2008).  In addition, the absorption
profiles reveal complex gas kinematics with $8-15$ individual
components per sightline over a line-of-sight velocity interval of
$\Delta\,v \approx 300-600$ \kms.

The observed relative abundance of singly-ionized iron and magnesium
has two important utilities. First of all, because $\mathrm{Mg^+}$ and
$\mathrm{Fe^+}$ share similar ionization potentials, the relative ionization correction is small in moderately ionized gas. As demonstrated in Zahedy \etal\ 2016, in most conditions the observed ionic column density ratio $\mathrm{\mathit{N}(Fe\,II)/\mathit{N}(Mg\,II)}$ places a lower limit on the underlying (Fe/Mg) elemental abundance ratio. Secondly, while iron is produced in both core collapse and Type Ia supernovae (SNe) (e.g., Tsujimoto \etal\ 1995), magnesium is an $\alpha$ element produced primarily in massive stars and core-collapse SNe (e.g., Nomoto \etal\ 2006).  The lower limit of the underlying Fe/Mg elemental abundance ratio from the observed $\mathrm{\mathit{N}(Fe\,II)/\mathit{N}(Mg\,II)}$ therefore provides a
powerful means of distinguishing between different chemical enrichment
sources.  Zahedy \etal\ (2016) found that the cool gas in the
vicinities of massive lensing galaxies exhibits a uniformly
super-solar $\mathrm{(Fe/Mg)}$ abundance pattern, and concluded that
the Fe-rich gas is most likely located in the interstellar medium
(ISM) of the lensing galaxies with a significant contribution ($\apg20
\%$) of Type Ia supernovae (SNe Ia) to the ISM chemical enrichment
history.

The findings of Zahedy \etal\ (2016) illustrate the potential of
applying the observed $\mathrm{\mathit{N}(Fe\,II)/\mathit{N}(Mg\,II)}$
to gain new insights into the origin of chemically-enriched diffuse
gas around distant galaxies.  Specifically, recently accreted gas from
the intergalactic medium (IGM) is expected to show $\alpha$-element
enhanced abundance pattern, reflecting its early enrichment history
(e.g., Rauch \etal\ 1997), while stripped gas from the ISM of evolved
satellites is expected to display a relatively more Fe-rich pattern
owing to a longer lifetime over which more SNe Ia can occur and
contribute a higher fraction of Fe-peak elements to the environment.

On the other hand, the Zahedy \etal\ (2016) study also raises an
interesting question regarding the extent of SNe Ia-dominated feedback
in quiescent galaxies.  It has been found that the radial distribution
of the rate of Type Ia SNe in early-type galaxies appears to follow
the same S\'ersic profile describing the stellar light distribution of
the galaxies (F\"{o}rster \& Schawinski 2008). If cool gas around
passive galaxies is locally enriched by SNe~Ia, then a radial
dependence in Fe/$\alpha$ should be expected, with a higher fraction
of Fe-rich absorbers observed at smaller $d$.  In this paper, we
expand the lensing sample studied in Zahedy \etal\ (2016), which
probes the ISM of massive quiescent galaxies at $d<15$ kpc, to include
both quiescent and star-forming galaxies with measurements of
$\mathrm{\mathit{N}(Fe\,II)}$ and $\mathrm{\mathit{N}(Mg\,II)}$
available out to $z\approx 200$ kpc.  This expanded sample allows us
to examine the radial profiles of the gas-phase Fe/$\alpha$ ratio in
galaxy halos based on the observed
$\mathrm{\mathit{N}(Fe\,II)/\mathit{N}(Mg\,II)}$.

This paper is organized as follows.  Section 2 presents the QSO-galaxy
pair sample, as well as the QSO spectroscopic data and corresponding
data reduction.  The absorption-line measurements are presented in
Section 3, and the observed absorption-line properties as a function
of projected distance from passive and star-forming galaxies are
presented in Section 4.  Finally, a discussion of the implications is
presented in Section 5.  Throughout this paper, a $\Lambda$ cosmology
of $\Omega_{\rm M}=0.3$ and $\Omega_\Lambda = 0.7$, with a Hubble
constant of $H_0 = 70 \ {\rm km} \ {\rm s}^{-1}\ {\rm Mpc}^{-1}$ is
adopted.

\section[]{Observational Data}

A sample of 27 intermediate-redshift galaxies is assembled for
investigating the radial profiles of the gas-phase Fe/$\alpha$ ratio
in galaxy halos.  This galaxy sample comprises 13 passive galaxies at
$z=0.40-0.73$ and 14 star-forming galaxies at $z=0.10-1.24$ from a
combination of a literature search and our own observations.  These
galaxies are probed by background QSO sightlines over a range of
projected distances, from $d\approx 3$ kpc to $d\approx 400$ kpc for
the subsample of passive galaxies and from $d\approx 8$ kpc to
$d\approx 160$ kpc for the subsample of star-forming galaxies.  Here
we describe the assembly of the galaxy sample, and associated echelle
spectroscopy of the background QSOs.

\begin{center}
\begin{table*}
  \centering
    \caption{Summary of galaxy and absorption properties}
\resizebox{7.in}{!}{
    \begin{tabular}{@{\extracolsep{4pt}}lcclcccccccr@{}}
      \hline
      %\multicolumn{1}{c}{QSO}& Galaxy &&  & & & &{$d$} &$W_r\mathrm{(2796)}$ & &   \\ \cline{10-11}
      %\multicolumn{1}{c}{Sightline}&ID&RA (J2000)& Dec (J2000)&   \multicolumn{1}{c}{$z_\mathrm{gal}$} &Ref$^a$ & Type$^b$ &\multicolumn{1}{c}{(kpc)} & \multicolumn{1}{c}{(\AA)}  & \multicolumn{1}{c}{log\, $\mathit{N}_{tot}$(Fe\,II)} & \multicolumn{1}{c}{log\, $\mathit{N}_{tot}$(Mg\,II)}  & \multicolumn{1}{r}{$\mathrm{log\,\langle\frac{ \mathit{N}(Fe\,II)}{\mathit{N}(Mg\,II)}\rangle}$}   \\ 
       \multicolumn{1}{c}{QSO}& \multicolumn{6}{c}{Galaxy}& \multicolumn{4}{c}{Absorption$^a$}   \\ \cline{2-7} \cline{8-12}
      \multicolumn{1}{c}{Sightline}&RA (J2000)& Dec (J2000)&   \multicolumn{1}{c}{$z_\mathrm{gal}$} &Ref.$^b$ &$L_B/L_*$ &\multicolumn{1}{c}{$d$ (kpc)} & \multicolumn{1}{c}{$W_r\mathrm{(2796)}$ (\AA)}  & \multicolumn{1}{c}{log\, $\mathit{N}_{tot}$(Fe\,II)$^c$} & \multicolumn{1}{c}{log\, $\mathit{N}_{tot}$(Mg\,II)$^d$} & \multicolumn{1}{c}{log\, $\mathit{N}_{tot}$(Mg\,I)$^c$}  & \multicolumn{1}{r}{$\mathrm{log\,\langle\frac{ \mathit{N}(Fe\,II)}{\mathit{N}(Mg\,II)}\rangle}$}   \\ 
      \multicolumn{1}{c}{(1)}&\multicolumn{1}{c}{(2)}& \multicolumn{1}{c}{(3)}&  \multicolumn{1}{c}{(4)}&\multicolumn{1}{c}{(5)}&\multicolumn{1}{c}{(6)}&\multicolumn{1}{c}{(7)}& \multicolumn{1}{c}{(8)}&\multicolumn{1}{c}{(9)} & \multicolumn{1}{c}{(10)}&\multicolumn{1}{c}{(11)}&\multicolumn{1}{c}{(12)}   \\ 

      \hline
      \hline
      \multicolumn{12}{c}{Passive galaxies}\\
      \hline
     HE\,0047$-$1756 {\it A} 	 				& 00:50:27.84 & $-$17:40:09.6  & 0.408 & Z16 & 0.9& 5 & $4.46\pm0.02$ & $14.61$ & $14.67$ &$12.97$ &$-0.06\pm0.14$ \\
     HE\,0047$-$1756 {\it B}	  				& 			 & 			 & 	     &        &        & 3 & $3.69\pm0.04$ & $14.70$ & $14.98$ &$12.86$ &$-0.28\pm0.16$ \\
     SDSSJ1025$+$0349 		   				& 10:25:32.53 &  $+$03:49:25.0 & 0.4650 & BOSS &  1.7&166 & $2.38\pm0.03$ & $14.45$ & $>14.48$&$12.41$ &$<-0.03$\\  
     HE\,1104$-$1805 {\it A}	  				& 11:06:33.45 & $-$18:21:24.5  & 0.729 & L00 &  3.5&8 & $0.64\pm0.01$ & $13.50$ & $13.44$ &$11.75$ &$0.06\pm0.04$ \\
     SDSSJ1144$+$0714 		   				& 11:44:45.30 &  $+$07:14:56.5 & 0.4906 & GC11 &  2.1&100 & $0.74\pm0.01$ & $14.03$ & $14.72$ &$12.41$ & $-0.69\pm0.16$\\
     SDSSJ1146$+$0207 		   				& 11:46:58.64 &  $+$02:07:16.8 & 0.5437 & GC11&  2.6&74 & $1.48\pm0.05$ & $14.45$ & $>14.23$&$12.34$ & $<0.22$\\  
     Q1241$+$176$^e$			  			& 12:44:09.17 &  $+$17:21:11.9 & 0.5591 & BOSS &  3.8&159& $0.14\pm0.01$ & $<11.80$ & $12.50$&$<10.61$ & $<-0.70$\\ 	
     & 12:44:11.05 &  $+$17:21:04.9 & 0.5507 & CSV96 &  0.5&21& $0.48\pm0.01$ & $13.68$ & $13.68$ &$12.00$ & $0.00\pm0.05$\\   
     SDSSJ1422$+$0414 		   				& 14:22:42.63 &  $+$04:15:12.0 & 0.5512 & GC11 &  2.6&211 & $0.28\pm0.01$ & $13.13$ & $13.71$ &$11.07$& $-0.58\pm0.04$\\
     SDSSJ1506$+$0419 		   				& 15:06:38.26 &  $+$04:19:06.9 & 0.6155 & GC11 &  3.4&299 & $0.35\pm0.02$ & $13.23$ & $13.69$&$<11.28$ & $-0.46\pm0.08$\\  
     3C\,336			 	   				& 16:24:37.52 &  $+$23:45:06.0 & 0.3675 & S97 &  1.2&113 & $0.25\pm0.01$ & $12.99$ & $13.11$&$11.13$ & $-0.12\pm0.22$\\  
     & 16:24:38.59 &  $+$23:45:21.9 & 0.3181 & S97 &  0.6&55 & $0.51\pm0.01$ & $13.39$ & $13.66$&$11.33$ & $-0.27\pm0.19$\\  
     SDSSJ2116$-$0624 		   				& 21:16:25.96 &  $-$06:24:15.4 & 0.5237 & GC11&  2.3&144 & $0.55\pm0.01$ & $13.45$ & $>13.80$&$11.66$ & $<-0.35$\\  
     SDSSJ2324$-$0951 		   				& 23:24:50.18 &  $-$09:50:48.5 & 0.5446 & GC11 &  2.1&398 & $1.65\pm0.03$ & $13.79$ & $>14.12$&$12.28$ & $<-0.33$\\  \hline
      \multicolumn{12}{c}{Star-forming galaxies}\\
      \hline

     PKS\,0122$-$0021 		   				& 01:25:27.67 &  $-$00:05:31.4 & 0.3985 & M15 & 1.8&163 & $0.34\pm0.02$ & $12.74$ & $14.29$&$11.92$ & $-1.55\pm0.28$\\  
     & 01:25:28.21 &  $-$00:05:54.6 & 0.9541 & M15 & 1.6  &77 & $0.12\pm0.01$ & $12.00$ & $12.91$&$<10.89$ & $-0.91\pm0.12$\\
     PKS\,0349$-$1438 		   				& 03:51:27.80 &  $-$14:28:58.2 & 0.3567 & C98 &  0.6&72 & $0.21\pm0.01$ & $13.07$ & $13.97$&$11.04$ & $-0.91\pm0.11$\\
     PKS 0439$-$433  		  				& 04:41:17.25 &  $-$43:13:40.1  & 0.1010 & CKR05 & 1.0  &8     & ... & $14.92$ & $15.51^f$&... & $-0.59\pm0.07^h$\\  
     PKS\,0454$-$2203 		   				& 04:56:08.82 &  $-$21:59:27.0 & 0.4838 & K10 &  1.4&108& $0.42\pm0.01$ & $13.49$ & $13.66$&$11.83$ & $-0.17\pm0.08$\\  
     Q\,1148$+$387$^e$	 		   				& 11:51:29.26 &  $+$38:25:56.4 & 0.5536 & S02 &  1.0&23 & $0.64\pm0.01$ & $13.49$ & $13.74$&$<11.08$ & $-0.25\pm0.05$\\  
     PKS\,1354$+$1933 		  				& 13:57:04.53 &  $+$19:19:15.1 & 0.4592 & EGY91 &  0.5&44& $0.85\pm0.01$ & $13.64$ & $13.88$&$11.86$ & $-0.24\pm0.14$\\  
     PKS\,1424$-$1150 		   				& 14:27:38.18 &  $-$12:03:32.9 & 0.3404 & C01 &  0.6&85& $0.15\pm0.02$ & $<12.73$ & $12.70$&$11.49$ & $<0.03$\\  
     SDSSJ1430$+$0149 		   				& 14:30:40.71 &  $+$01:49:40.8 & 1.2418 & Z16b & 0.8  &17 & $2.86\pm0.01$ & $15.23$ & $15.48^g$&$13.04$ & $-0.25\pm0.09^i$\\
     3C\,336 				   				& 16:24:38.19 &  $+$23:45:22.0 & 0.702 & S97 &  0.5&113 & $0.040\pm0.003$& $<11.35$& $12.16$&$<10.25$ & $<-0.81$\\
     						   			& 16:24:38.42 &  $+$23:45:15.2 & 0.798 & S97 &  0.4& 71 & $0.45\pm0.01$ & $12.59$ & $13.29$& $<11.02$ & $-0.70\pm0.08$\\
     & 16:24:38.77 &  $+$23:45:08.3 & 0.472 & S97 &  0.2& 34 & $0.81\pm0.01$ & $13.52$ & $14.12$&$11.70$ & $-0.59\pm0.12$\\  
     & 16:24:39.30 &  $+$23:45:12.1 & 0.892 & S97 &  0.7&  23& $1.55\pm0.01$ & $14.73$ & $>14.50$&$12.50$ & $<0.23$\\
     Q2206$-$199 			   				& 22:08:51.55 &  $-$19:43:52.6 & 0.948 & GB97 &  2.1&87 & $0.25\pm0.01$ & $12.48$ & $13.37$& $10.87$ & $-0.89\pm0.15$\\
      \hline
      \multicolumn{12}{l}{\bf Notes} \\
      \multicolumn{12}{l}{$^a$\ The typical uncertainty in total column densities of Fe\,II, Mg\,II, and Mg\,I absorption (Columns 9 through 11) is smaller than 0.05 dex for all absorption systems; see \S\ 3.}\\
      \multicolumn{12}{l}{\ \ The quoted uncertainty in $\mathrm{log\,\langle \mathit{N}(Fe\,II)/\mathit{N}(Mg\,II)\rangle}$ (Column 12) for each system is the dispersion from the weighted mean, which represents the scatter in $\mathrm{\mathit{N}(Fe\,II)/\mathit{N}(Mg\,II)}$}\\
      \multicolumn{12}{l}{\ \ among individual components of the corresponding absorber.}\\
      \multicolumn{12}{l}{$^b$\ References for galaxy redshifts: Ellingson, Green \& Yee 1991 (EGY91); Churchill, Steidel \& Vogt 1996  (CSV96); Steidel \etal\ 1997 (S97); Guillemin \& Bergeron}\\
     \multicolumn{12}{l}{\ \ 1997 (GB97); Chen \etal\ 1998 (C98); Lidman \etal\ 2000 (L00); Chen \etal\ 2001 (C01); Steidel \etal\ 2002 (S02); Chen, Kennicutt \& Rauch 2005 (CKR05);}\\ 
     \multicolumn{12}{l}{\ \ Kacprzak \etal\ 2010 (K10); Gauthier \& Chen 2011 (GC11); Dawson \etal\ 2013 (BOSS); Muzahid \etal\ 2015 (M15);  Zahedy \etal\ 2016 (Z16); This work (Z16b).}\\      
	
     \multicolumn{12}{l}{$^c$ The upper limits are  2-$\sigma$ limits on the total column density for non detections, estimated using
       the error spectrum.}  \\

     \multicolumn{12}{l}{$^d$ The lower limits represent  2-$\sigma$ limits on the total column density for saturated Mg\,II $\lambda\lambda\ 2796,2803$ absorption doublet.}\\

     \multicolumn{12}{l}{$^e$ Absorption-line measurements for these fields are adopted from Churchill \etal\ (2000) and Churchill \& Vogt (2001).}\\

     \multicolumn{12}{l}{$^f$ Because the Mg\,II $\lambda
       \lambda\ 2796,2803$ absorption doublet falls outside the
       observable spectral window for this QSO sightline, the total
       $N$(Mg\,II) of the system was estimated}\\

     \multicolumn{12}{l}{\ \ from the total S\,II column density of the
       system, log\,$N_\mathrm{tot}$(S\,II); see \S\ 3.}\\

     \multicolumn{12}{l}{$^g$\ Because the Mg\,II $\lambda
       \lambda\ 2796,2803$ absorption doublet is heavily saturated for
       this absorption system, log\,$N_\mathrm{tot}$(Mg\,II) was
       estimated from the total Si\,II column density}\\

     \multicolumn{12}{l}{\ \ of this system,
       log\,$N_\mathrm{tot}$(Si\,II); see \S\ 3.}\\

     \multicolumn{12}{l}{$^h$ The error is estimated by propagating
     the statistical errors for log\,$\mathit{N}_{tot}$ (Fe,II), log\,$\mathit{N}_{tot}$(S\,II), and the uncertainty in the adopted (Mg/S) ratio from Asplund \etal\ 2009; see \S\ 3.}\\

     \multicolumn{12}{l}{$^i$\ The error is estimated by propagating
     the statistical errors for log\,$\mathit{N}_{tot}$ (Fe,II), log\,$\mathit{N}_{tot}$(Si\,II), and the uncertainty in the adopted (Mg/Si) ratio from Asplund \etal\ 2009; see \S\ 3.}\\

   \label{gal_table}
  \end{tabular}
}
\end{table*}
\end{center}

\subsection[]{Absorbing Galaxy Sample}

We first performed a literature search of intermediate-redshift
galaxies that are associated with known Mg\,II absorbers and have
high-resolution echelle absorption spectra of the background QSOs
available in the public data archives.  Our search resulted in 16
Mg\,II absorbing galaxies at redshifts $0.10\leq z \leq 0.95$ in 10
QSO fields: PKS\,0122$-$0021, PKS\,0349$-1438$, PKS\,0439$-$433,
PKS\,0454$-$2203, Q1148$+$387, Q1241$+$176, PKS\,1354$+$1933,
PKS\,1424$-$1150, 3C\,336, and Q2206$-$199.  A number of studies have
previously investigated the photometric and spectroscopic properties
of these absorbing galaxies (e.g, Steidel \etal\ 1997, 2002; Chen
\etal\ 1998, 2001, 2005, 2008; Kacprzak \etal\ 2010, 2011), based on
high-resolution {\it Hubble Space Telescope} ({\it HST}) imaging and
deep ground-based spectroscopic observations.  We utilized these
studies to classify the selected galaxies into two subsamples:
star-forming and passive galaxies.  Star-forming galaxies are
classified based on an emission-line dominated spectrum or a
disk-dominated light profile (disk-to-bulge light ratios $>3$, e.g.,
Chen \etal\ 1998) when the galaxy spectrum is not available.  Passive
galaxies are classified based on an absorption-line dominated spectrum
and a bulge-dominated light profile (disk-to-bulge light ratios of
$<3$).  Following these criteria, 13 galaxies are classified as
star-forming galaxies, whereas 3 galaxies are classified as passive
galaxies.  In addition to these previously known Mg\,II-absorbing
galaxies, we included a newly identified $z_\mathrm{gal}=1.2418$
star-forming galaxy at $d=17$ kpc from the QSO SDSSJ\,1430$+$0149,
where an ultra-strong Mg\,II absorber of rest-frame absorption
equivalent width $W_r(2796)\approx 2.8$ \AA\ had been identified at
the same redshift (e.g., Zych \etal\ 2009).

To increase the number of passive galaxies in our sample, we included
Luminous Red Galaxies (LRGs) at $z\sim0.5$ with associated Mg\,II
absorption features from Gauthier \& Chen (2011).  We further
supplemented this passive galaxy subsample with new, unpublished
Mg\,II-absorbing LRGs from our own survey.  Together with two massive
lensing galaxies with associated Mg\,II absorption from Zahedy
\etal\ (2016), this process resulted in 10 additional passive galaxies
probed by 11 QSO sightlines at $d=3-398$ kpc.  A summary of the
properties of the final sample of 27 galaxies is presented in Columns
2 to 7 of Table 1, where we list for each galaxy its right ascension
and declination, redshift $z_\mathrm{gal}$, galaxy projected distance
from the QSO sightline $d$, and the $B$-band luminosity $L_B$ in unit
of $L_*$, calculated using the redshift-dependent absolute $B$-band
magnitude measurements from Faber \etal\ (2007).

\subsection[]{QSO Absorption Spectroscopy}

High-resolution echelle spectra of the QSOs PKS\,0349$-$1438,
PKS\,1424$-$1150, SDSS\,J1430$+$0149, 3C336, and Q2206$-$199 were
obtained using the Ultraviolet and Visual Echelle Spectrograph (UVES;
D'Odorico et al.\ 2000) on the VLT-UT2 telescope under multiple
observing programs (PIDs 076.A-0860(A), 075.A-0841(A), 079.A-0656(A),
081.A-0478(A), 69.A-0371(A), and 65.O-0158(A), respectively).  We
retrieved the reduced and order-combined individual exposures from the
ESO Advanced Data Products Archive. Following data retrieval, we
performed vacuum and heliocentric corrections to the QSO spectra,
co-added different exposures, and performed continuum fitting to the
data. The resulting spectra typically have a high signal-to-noise
ratio of $S/N>15$ per resolution element of $7-8$ \kms\ in
full-width-at-half-maximum (FWHM).

Echelle spectroscopic observations were obtained using the MIKE
echelle spectrograph (Bernstein \etal\ 2003) on the Magellan Clay
Telescope for the QSOs PKS\,0122$-$0021, PKS\,0349$-1438$,
PKS\,0454$-$2203, SDSSJ1144$+$0714, SDSSJ1422$+$0414,
SDSSJ1506$+$0419, SDSSJ1025$+$0349, SDSSJ1146$+$0207,
SDSSJ2116$-$0624, and SDSSJ2324$-$0951 during multiple observing runs
between October 2007 and April 2015.  A $1''$ slit and $2\times2$
binning were employed for the observations of PKS\,0122$-$0021,
PKS\,0349$-1438$, PKS\,0454$-$2203, SDSSJ1144$+$0714,
SDSSJ1422$+$0414, SDSSJ1506$+$0419, SDSSJ2116$-$0624, and
SDSSJ2324$-$0951.  A $0.7''$ slit and $2\times2$ binning were employed
for the observations of SDSSJ1146$+$0207, and a $0.7''$ slit and
$3\times3$ binning were employed for the observations of
SDSSJ1025$+$0349.  MIKE delivers spectral resolutions of ${\rm
  FWHM}\sim 12$ \kms\ and 8 \kms\ for the $1.0''$ and $0.7''$ slits,
respectively.  The spectra were reduced using a custom data reduction
pipeline previously described in Chen \etal\ (2014) and Zahedy
\etal\ (2016).  The final combined spectra are characterized by
$S/N>10$ per resolution element at $\lambda>3500$ \AA.  Finally,
details of the data reduction for the echelle spectra of the two
lensed QSO systems HE\,0047$-$1756 (MIKE) and HE\,1104$-$1805 (UVES and Keck HIRES) have previously
been described in Zahedy \etal\ (2016).  A journal of the echelle
spectroscopic observations of the QSO sightlines in our study is shown
in Table 2.

\begin{table}
\begin{center}
\caption{Journal of QSO echelle spectroscopy}
\vspace{-0.5em}
\label{tab:MIKE,HIRES}
\resizebox{3.25in}{!}{
\begin{tabular}{lcccl}\hline
\multicolumn{1}{c}{QSO Image}    &$z_{\rm{em}}$ & Instrument 	& \multicolumn{1}{c}{Exp. time (s)} 	& \multicolumn{1}{c}{Date} \\	
% 		&  	&         		& \multicolumn{1}{c}{(s)}         &        \\
\multicolumn{1}{c}{(1)}		&\multicolumn{1}{c}{(2)}  	&\multicolumn{1}{c}{(3)}&\multicolumn{1}{c}{(4)}& \multicolumn{1}{c}{(5)}        \\
\hline \hline		
 3C336				& 0.927	& UVES 	&     9800		& 2002 Apr, May \\  % 69.A-0371(A) Savaglio
HE\,0047$-$1756$A$ 	& 1.676	& MIKE	& 	7200		& 2013 Nov \\
HE\,0047$-$1756$B$	& 1.676	& MIKE	&     5700		& 2013 Nov \\  
HE\,1104$-$1805$A$	& 2.305	& HIRES	&     19300	& 1997 Feb \\ 
					& 		& UVES 	&     19000	& 2001 Jun \\ 		
 PKS\,0122$-$0021		& 1.077	& MIKE	&	1200		& 2007 Oct\\
 PKS\,0349$-$1438 		& 0.616	& MIKE	&	1800		& 2007 Oct \\
 PKS\,0454$-$2203 		& 0.533	& MIKE	&	2700		& 2007 Oct \\
 PKS\,1354$+$1933		& 0.720 	& UVES 	&      600		& 2001 Jun \\  %076.A-0860(A) Miniati
 PKS\,1424$-$1150 		& 0.806 	& UVES 	&      720		& 2005 Jul \\  %075.A-0841(A) Miniati
 Q2206$-$199 			& 2.558 	& UVES 	&    17100		& 2000 May \\  % 65.O-0158(A) Pettini
SDSSJ1025$+$0349 	& 1.325	& MIKE	&	5400		& 2015 Apr \\		
 SDSSJ1144$+$0714 	& 0.919	& MIKE	&	6000		& 2010 May \\		
 SDSSJ1146$+$0207	& 1.137	& MIKE	&	9000		& 2010 May \\ 		
 SDSSJ1422$+$0414 	& 0.972	& MIKE	&	6000		& 2010 May \\		
 SDSSJ1430$+$0149	& 2.119    & UVES 	&    12150		& 2007 May, Jun \\  %079.A-0656(A) Zych
					&		& UVES 	&    2930		& 2008 Aug \\  %081.A-0478(A) Zych
 SDSSJ1506$+$0419 	& 1.288	& MIKE	&	9000		& 2010 May \\		 
 SDSSJ2116$-$0624 	& 1.042	& MIKE	&	8000		& 2010 May \\		 
 SDSSJ2324$-$0951	& 0.764	& MIKE	&	9000		& 2010 Sep \\ 		
\hline
\end{tabular}}
\end{center}
\end{table}

\section[]{Absorption Line Analysis}

The high-resolution echelle spectra of the QSOs described in \S\ 2 enable accurate constraints on both the integrated rest-frame Mg\,II absorption equivalent width ($W_r\mathrm{(2796)}$) and the ionic column densities of Fe$^+$, Mg$^+$, and Mg$^0$ for each galaxy in our sample.  In particular, a component-by-component analysis allows us to
examine how the relative abundance ratios between any two ions vary
within individual galaxy halos and across the full sample (e.g.,
Zahedy \etal\ 2016).  We employ a custom software, previously
developed by and described in Zahedy \etal\ (2016), to perform a Voigt
profile fitting analysis for constraining the Mg\,II, Mg\,I, and
Fe\,II column densities of individual absorbing components.  We
perform a simultaneous fit to prominent absorption transitions,
including the Mg\,II\,$\lambda\lambda\,2796, 2803$ doublet,
Mg\,I\,$\lambda\,2852$, and a series of Fe\,II transitions.  For all
galaxies, Fe\,II\,$\lambda\,2600$ and Fe\,II\,$\lambda\,2586$
transitions are included in the Voigt profile analysis.  For galaxies
at $z\apg 0.5$, additional Fe\,II\,$\lambda\,2382$,
Fe\,II\,$\lambda\,2374$, and Fe\,II\,$\lambda\,2344$ transitions are included.

For individual Mg\,II absorbing components with no corresponding
Fe\,II or Mg\,I absorption, we measure a 2-$\sigma$ upper limit in the
absorption equivalent width of the strongest transitions using the error
spectrum.  The upper limits are evaluated over a spectral window twice
the FWHM of the corresponding Mg\,II component.  The measured
2-$\sigma$ equivalent width limits are then converted to 2-$\sigma$
upper limits of the component column densities under an optically-thin
assumption.  For saturated Mg\,II absorbing components, we place
2-$\sigma$ lower limits of the component column densities based on a
grid search of the $\chi^2$ values from the Voigt profile fitting
results.

In five cases, we cannot measure the Mg\,II absorption strength
directly either due to missing echelle spectra in the public archives
or a lack of spectral coverage for the relevant transitions.  Three of
these systems are identified along QSO sightlines Q1241$+$176 and
Q1148$+$387.  The observed $W_r\mathrm{(2796)}$ and total Fe\,II,
Mg\,II, and Mg\,I column density measurements for these systems have been published in Churchill \etal\ (2000) and Churchill \& Vogt (2001).  We adopt these values for our subsequent analysis.  In addition, the galaxy at
$z=0.101$ in the field of PKS\,0439$-$433 does not have Mg\,II
absorption spectra available.  The Mg\,II absorber associated with the
galaxy at $z=1.24$ in the field of SDSSJ\,1430$+$0549 is heavily
saturated, and no useful constraint for the Mg\,II absorption column
density is available.  Finally, two lensing galaxies have been
published in Zahedy \etal\ (2016).  Excluding these galaxies leads to
20 galaxies for which we can perform our own Voigt profile analysis.
The results of the component-by-component Voigt profile analysis of
Mg\,II, Mg\,I, and Fe\,II absorption for these 20 galaxies are
presented in Figure 1.

\begin{figure*} 
%\hspace{-0.5em}
\includegraphics[width=176mm]{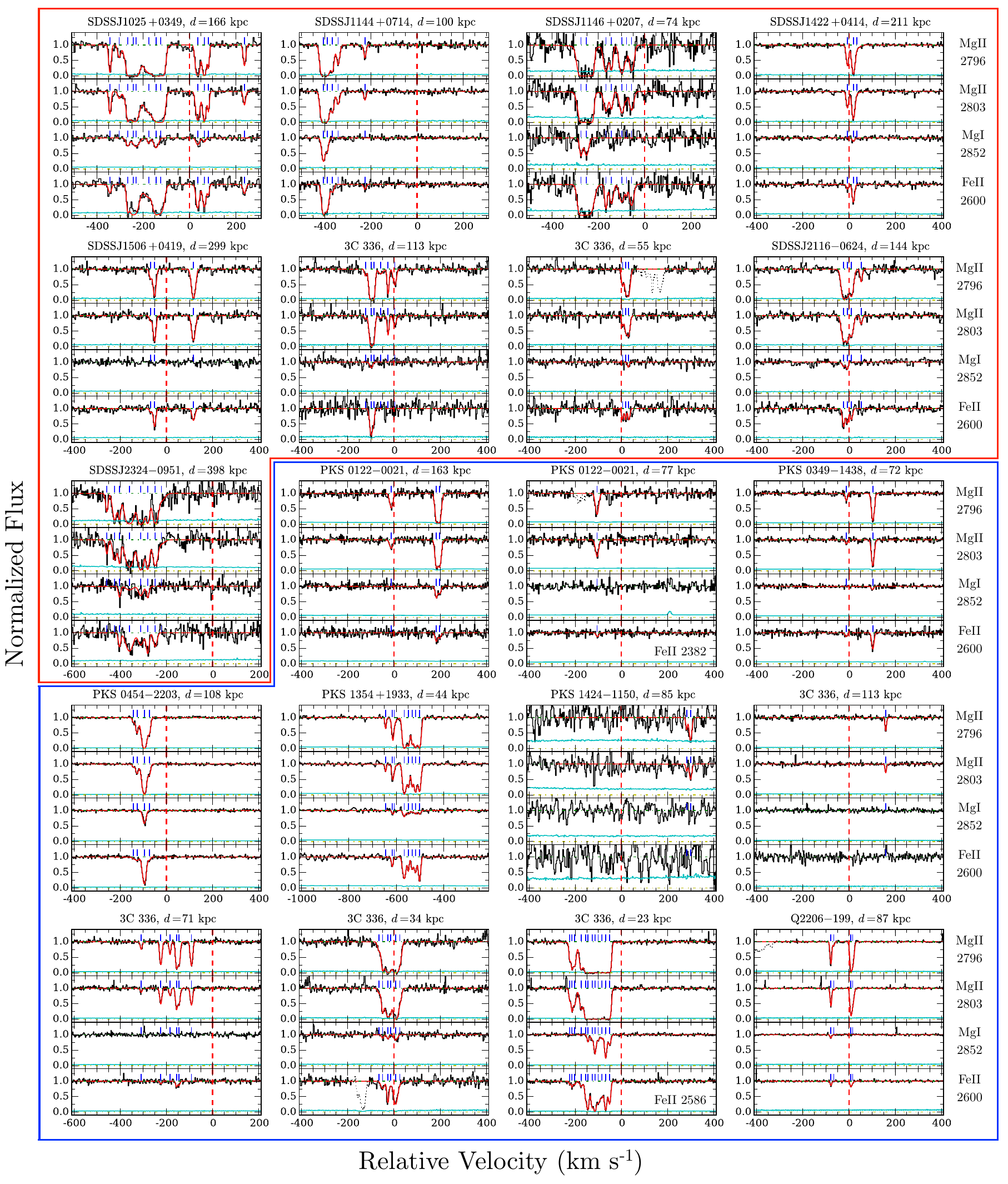}
%\vspace{-1.5em}
\caption{Summary of the component-by-component Voigt profile analysis
  of Mg\,II, Mg\,I, and Fe\,II absorption for 20 galaxies that have
  not been published previously.  For each galaxy, the observed
  absorption files of the Mg\,II doublet, Mg\,I\,$\lambda$\,2852, and
  Fe\,II\,$\lambda$\,2600 are presented from top to bottom panels.  In
  each panel, the absorption spectra and corresponding 1-$\sigma$
  error array are shown in black and cyan, respectively.  The best-fit
  Voigt profile is shown in red. Contaminating features have been dotted out
  for clarity.   The centroid of each absorbing
  component is marked by a blue tick mark at the top of each panel.
  Zero velocity corresponds to the systemic redshift of the absorbing
  galaxy.  Galaxy projected distance is indicated at the top of each
  four-panel block.  Finally, the red outlining box encloses passive
  galaxies and blue includes star-forming galaxies. }
\label{Figure 1}
\end{figure*}

For the galaxy at $z=0.101$ and $d=8$ kpc from PKS\,0439$-$433, a
damped Lyman \lya\ (DLA) absorption feature is found in the QSO
spectrum.  Measurement of the total Fe\,II column density,
$N_\mathrm{tot}$(Fe\,II), is based on far-ultraviolet transitions
observed using the Cosmic Origins Spectrograph (COS; Green
\etal\ 2012) on board {\it HST}.  We adopt
log\,$N_\mathrm{tot}$(Fe\,II)$=14.92\pm0.03$ from Som \etal\ (2015).
Furthermore, because no Mg\,II column density measurement is available
for this galaxy, we infer the total Mg\,II column density,
$N_\mathrm{tot}$(Mg\,II), from the reported total column density of another 
$\alpha$-element ion $\mathrm{S}^+$,  $N_\mathrm{tot}$(S\,II).  For this DLA, Som \etal\ (2015) measured log\,$N_\mathrm{tot}$(S\,II)$=15.03\pm0.03$.  We assume a solar elemental abundance pattern of $\log\,({\rm Mg}/{\rm S})_\odot=0.5$ dex (Asplund \etal\ 2009) for this system, motivated by the observed near-solar metallicity of the DLA (e.g., Chen \etal\ 2005; Som \etal\ 2015).  To investigate the ionization correction between the elemental ratio (Mg/S) and the observed ionic ratio $N_\mathrm{tot}$(Mg\,II)/$N_\mathrm{tot}$(S\,II), we perform photoionization calculations using \textsc{Cloudy} (Ferland \etal\ 2013; v.13.03) for a $T = 10^4$ K cloud with neutral hydrogen column density and metallicity reported for this system (log\,$N \mathrm{(H\,I)}=19.63$; [S/H]$=0.1$). We assume a plane-parallel
geometry for the gas cloud, which is illuminated on both sides with an
updated Haardt \& Madau (2001) ionizing radiation field (HM05 in
\textsc{Cloudy}) at $z = 0.1$. Using the photoionization model and
applying $N_\mathrm{tot}$(Fe\,II)/$N_\mathrm{tot}$(Fe\,III) of the
absorbing gas for constraining the mean ionization parameter, we find
that the ionization correction is negligible between (Mg/S) and
$N_\mathrm{tot}$(Mg\,II)/$N_\mathrm{tot}$(S\,II).  Based on the
assumed $({\rm Mg}/{\rm S})_\odot$, we infer
log\,$N_\mathrm{tot}$(Mg\,II)$=15.51\pm0.06$, where the error is estimated by propagating the statistical error for log\,$N_\mathrm{tot}$(S\,II) and the uncertainty in the adopted (Mg/S) ratio from Asplund \etal\ 2009. 

\begin{figure} 
\hspace{-0.5em}
\includegraphics[width=86mm]{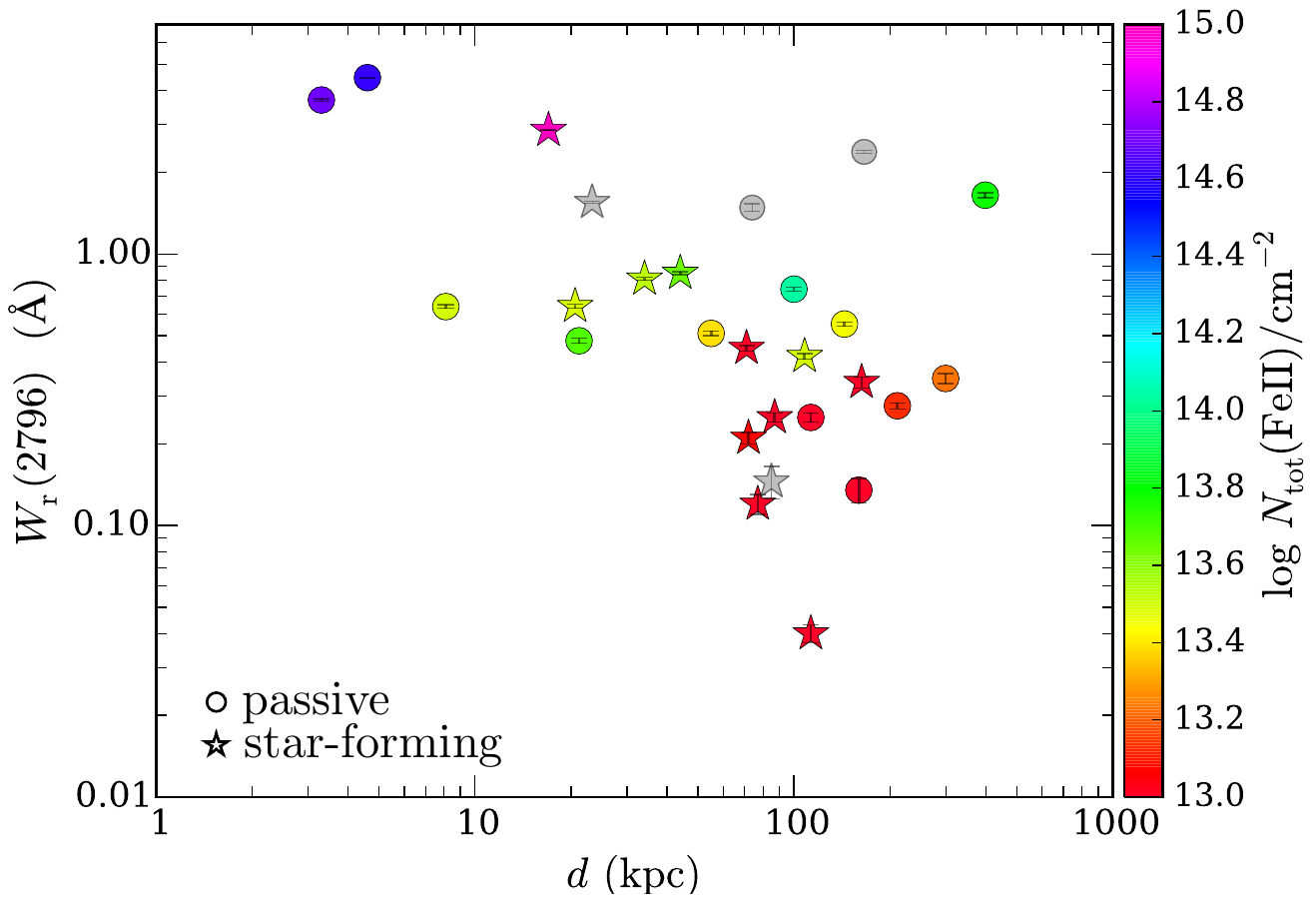}
\vspace{-1.5em}
\caption{Rest-frame Mg\,II absorption equivalent width $W_r$(2796)
  versus galaxy projected distance $d$ for the galaxy sample in this
  study.  Circles represent passive galaxies, while stars represent
  star-forming galaxies.  The color of each data point represents the
  total Fe\,II column density, log\,$N_\mathrm{tot}$(Fe\,II) of the
  absorption system associated with each galaxy.  Greyed out data
  points mark absorption systems with no constraints on the mean
  Fe\,II/Mg\,II ratio (see Figure 4 below) due to saturation (three
  galaxies) and poor signal-to-noise ratio (one galaxy).  We note that
  the Mg\,II doublet falls outside the observable window for PKS
  0439$-$433 and thus is not plotted here (see also \S\ 3 for a
  detailed discussion on this system).}
\label{Figure 2}
\end{figure}

\begin{figure} 
\hspace{-0.5em}
\includegraphics[width=87mm]{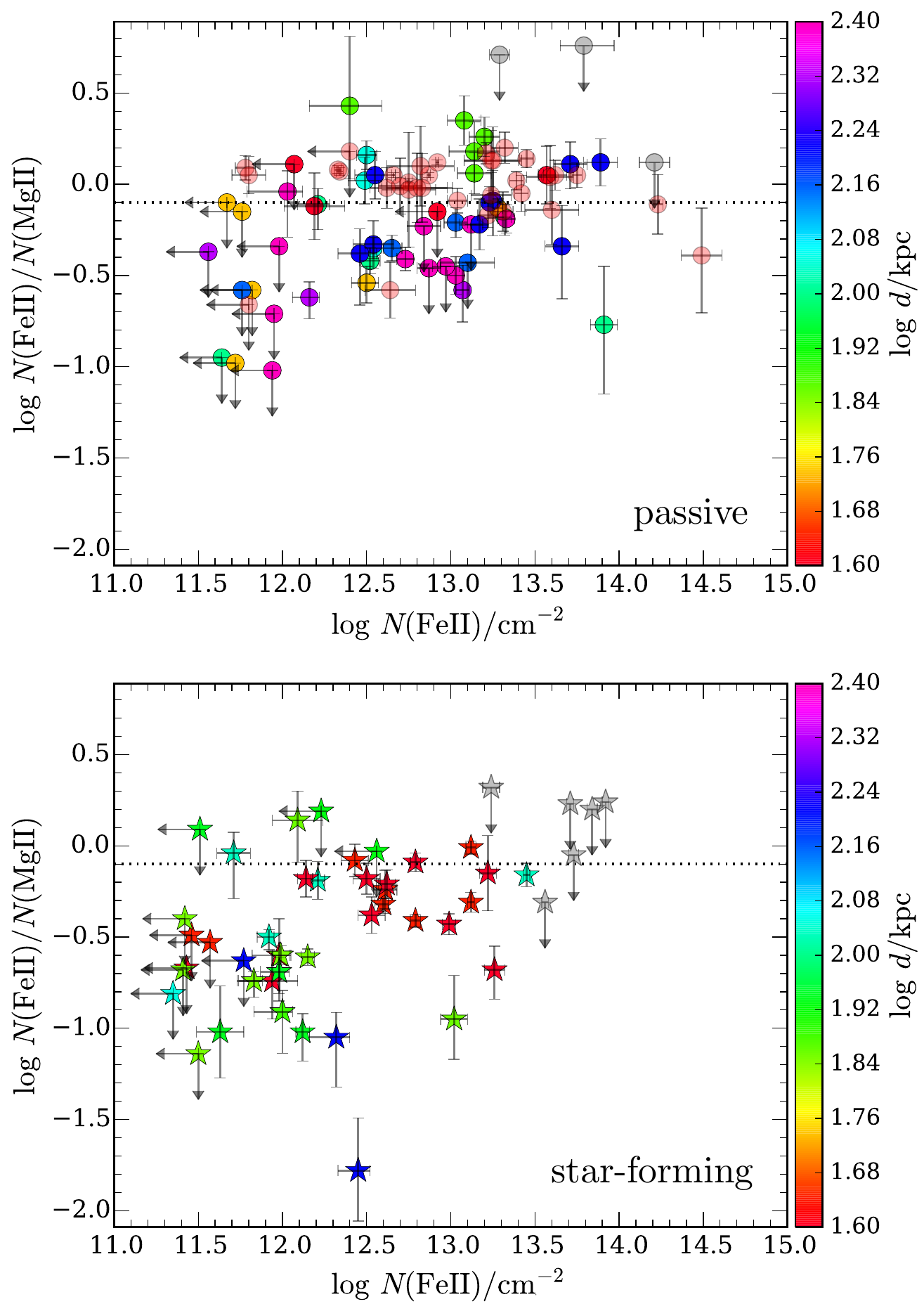}
\vspace{-1.5em}
\caption{Observed Fe\,II to Mg\,II column density ratios versus Fe\,II
  column density for individual components associated with passive
  (circles in the top panel) and star-forming (star symbols in the
  bottom panel) galaxies in this study.  Error bars associated with
  each data point show the measurement uncertainties.  The color of
  each data point represents the projected distance $d$ of the
  absorbing galaxy.  Saturated components associated with galaxies
  with poor constraints on the mean Fe\,II/Mg\,II ratio (see Figure 4
  below) are greyed out.  Both lensing galaxies in the passive
  subsample exhibit strong Mg\,II absorption at $d<10$ kpc.  The
  absorbers are resolved into $8-15$ components and show
  pre-dominantly Fe-rich gas (Zahedy \etal\ 2016).  Components
  associated with these lensing galaxies are shown in pale red to be
  separated from non-lensing galaxies at $d>10$ kpc in the top panel.
  In both panels, when a Mg\,II absorbing component has no
  corresponding Fe\,II absorption detected, it is shown as downward
  and left-pointing arrows with the data point indicating the
  2-$\sigma$ upper limit of $N({\rm Fe\,II})$.  In addition, when
  Mg\,II components are saturated, the inferred 2-$\sigma$ lower limit
  on $N({\rm Mg}\,II)$ directly translates to a 2-$\sigma$ upper limit
  on $N({\rm Fe\,II})/N({\rm Mg\,II})$.  The dotted horizontal line
  indicates the solar (Fe/Mg) abundance pattern from Asplund
  \etal\ (2009), log (Fe/Mg)$_\odot  = -0.10$, to guide visual comparisons.}
\label{Figure 3}
\end{figure}

For the galaxy at $z=1.2418$ and $d=17$ kpc from SDSSJ1430$+$0149, the
Mg\,II absorption line is heavily saturated.  We infer
$N_\mathrm{tot}$(Mg\,II) from the absorption strength of the weaker
and non-saturated Si\,II\,$\lambda\,1808$ transition also recorded in
the public UVES spectra.  A Voigt profile analysis returns a best-fit
integrated log\,$N_\mathrm{tot}$(Si\,II)$=15.75\pm 0.05$, which is
consistent with the published value in Zych \etal\ (2009).  We adopt a
solar elemental abundance ratio of $\log\,({\rm Mg}/{\rm
  Si})_\odot=0.1$ dex for this system, which is consistent with what
has been observed for a number of $z>1$ DLAs (e.g., Dessauges-Zavadsky
\etal\ 2006).  The ionization correction between (Mg/Si) and ionic
ratio $N_\mathrm{tot}$(Mg\,II)/$N_\mathrm{tot}$(Si\,II) is estimated
using a \textsc{Cloudy} photoionization model for a plane-parallel
cloud with 0.1 solar metallicity and log\,$N \mathrm{(H\,I)}=19.5$,
which is illuminated on both sides with the HM05 radiation field at
$z=1.24$. Using the model output and the observed
$N_\mathrm{tot}$(Mg\,I)/$N_\mathrm{tot}$(Si\,II) ratio, we estimate
$N_\mathrm{tot}$(Mg\,II)/$N_\mathrm{tot}$(Si\,II)$=0.44\times N({\rm
  Mg})/N({\rm Si})$ for this system and infer
log\,$N_\mathrm{tot}$(Mg\,II)\,$=15.48\pm0.07$, where the error is estimated by propagating the statistical error for log\,$N_\mathrm{tot}$(Si\,II) and the uncertainty in the adopted (Mg/Si) ratio from Asplund \etal\ 2009. 

The results of the absorption-line analysis are summarized in Columns
8 through 11 of Table 1, which present $W_r\mathrm{(2796)}$ and the
total column densities of Fe\,II, Mg\,II, and Mg\,I absorption,
log\,$N_\mathrm{tot}$(Fe\,II), log\,$N_\mathrm{tot}$(Mg\,II),
log\,$N_\mathrm{tot}$(Mg\,I), summed over all individual components.
Upper limits of the total column densities indicate non-detections,
whereas lower limits indicate saturated absorption.  The typical
uncertainty in the total integrated column densities is smaller than
0.05 dex for all absorption systems.

The general absorption properties of the galaxy sample are summarized
by the radial profile of Mg\,II absorption in Figure 2, which displays
the observed $W_r\mathrm{(2796)}$ versus projected distance $d$ for
all galaxies in the sample where $W_r\mathrm{(2796)}$ measurements are available.  The data points are color-coded
according to %the observed total Fe\,II column density,
$N_\mathrm{tot}$(Fe\,II) associated with each galaxy.  Absorption
systems with poor constraints on the $N({\rm Fe\,II})/N({\rm Mg\,II)}$
ratio due to either saturation or low signal-to-noise ratios are shown
in gray (see below).  With the exception of two passive galaxies at
$d>100$ kpc which exhibit saturated Mg\,II absorption features, the
Mg\,II absorbing gas around the galaxies in our sample follow a
similar declining trend of $W_r\mathrm{(2796)}$ with increasing $d$ as
presented in Chen \etal\ (2010) and Nielsen \etal\ (2013).  In
addition, the observed $N({\rm Fe\,II})$ also appears to decline with
increasing $d$.

To examine the relative Fe to Mg abundance pattern, we present in
Figure 3 the Fe\,II to Mg\,II column density ratio, $N({\rm
  Fe\,II})/N({\rm Mg\,II})$, versus Fe\,II column density, $N({\rm
  Fe\,II})$, for individual absorbing components associated with
passive (circles in the top panel) and star-forming (star symbols in
the bottom panel) galaxies.  The data points are color-coded by the
projected distances of the absorbing galaxies, and components
associated with the two lensing galaxies (both passive) at $d<10$ kpc
are shown in pale red to be separated from absorbing components found
at $d>10$ kpc from non-lensing galaxies.  When Fe\,II absorption is
not detected or when Mg\,II components are saturated, a 2-$\sigma$
limit is placed on $N({\rm Fe\,II})/N({\rm Mg\,II})$.  For
heavily saturated Mg\,II components, no sensitive constraints can be
obtained.  These components are shown in grey.

Two interesting features are seen in Figure 3.  First, passive
galaxies (including both lensing and non-lensing galaxies) display a
large fraction ($>50$\%) of Fe\,II-rich gas at $d<100$ kpc (data
points with colors red through green) with log\,$N({\rm
  Fe\,II})/N({\rm Mg\,II})>-0.1$ dex.  At larger distances, $d\apg
100$ kpc, only five components associated with two passive galaxies
show predominantly Fe\,II-rich content and the remaining 33 components
are consistent with Mg\,II-rich gas with log\,$N({\rm Fe\,II})/N({\rm
  Mg\,II}) \apl-0.1$ dex (data points with colors blue through
magenta).  Secondly, the majority of absorbing components associated
with star-forming galaxies are consistent with an Mg\,II-rich content
with log\,$N({\rm Fe\,II})/N({\rm Mg\,II}) \apl-0.1$ dex over the full
projected distance range probed by the sample.  Only three out of 43
non-saturated components display log\,$N({\rm Fe\,II})/N({\rm Mg\,II})
\apg-0.1$ dex.

\begin{figure} 
\begin{center}
\includegraphics[width=86mm]{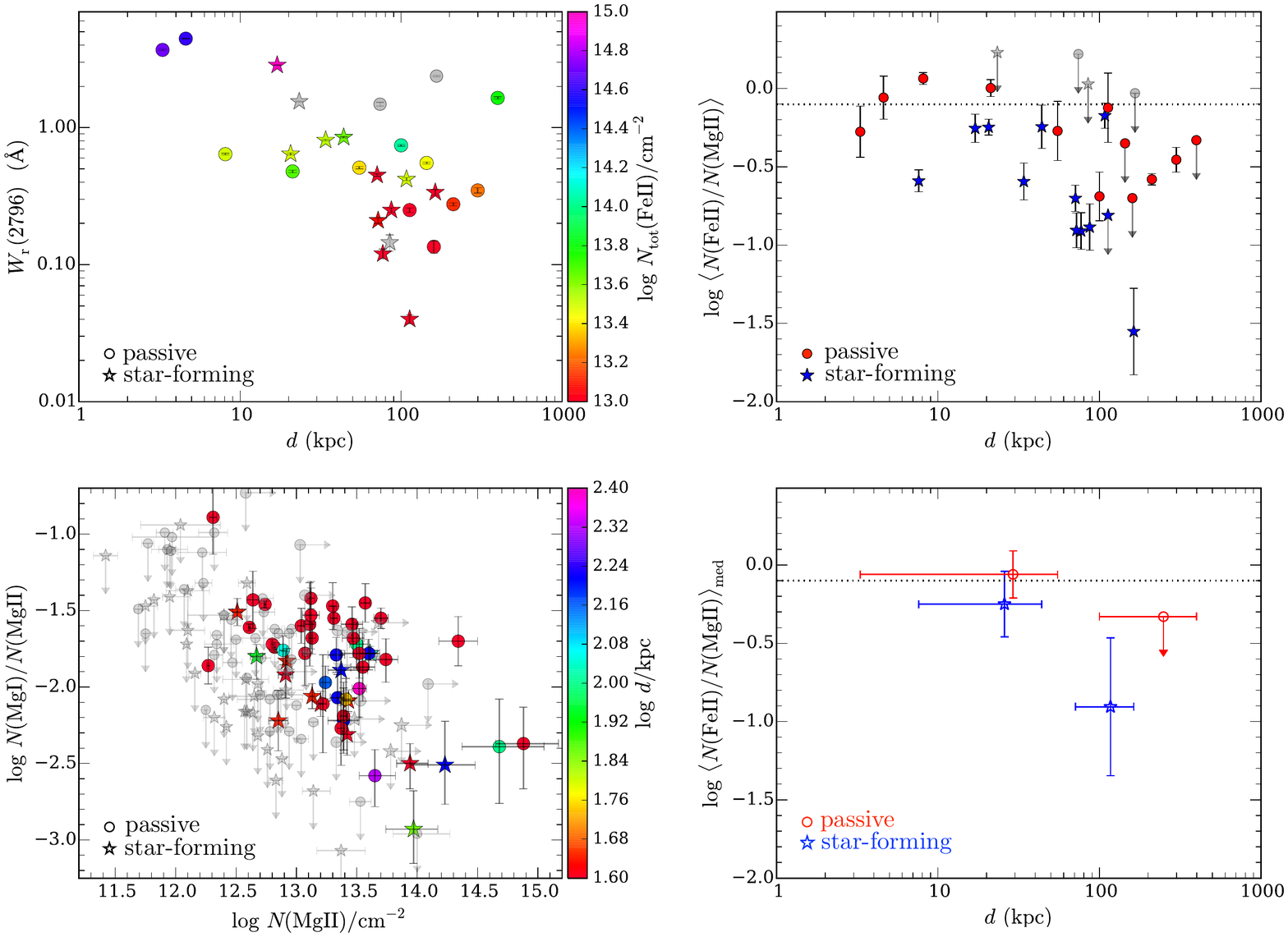}
\end{center}
\caption{Mean Fe\,II to Mg\,II column density ratio, $\mathrm{\langle
    \mathit{N}(Fe\,II)/\mathit{N}(Mg\,II)\rangle}$, from Equation (1)
  versus galaxy projected distance, $d$, for absorbers associated with
  passive galaxies (red circles) and for those associated with
  star-forming galaxies (blue stars).  Error bars for each data point
  represent the dispersion among individual component Fe\,II to Mg\,II
  ratios of the corresponding absorber.  As shown in Figure 3, only
  2-$\sigma$ upper limits can be placed on $N({\rm Fe\,II})/N({\rm
    Mg\,II})$, if Fe\,II absorption is not detected or if Mg\,II
  components are saturated.  These upper limits are propagated into
  the estimates of $\mathrm{\langle
    \mathit{N}(Fe\,II)/\mathit{N}(Mg\,II)\rangle}$ for each absorbing
  galaxy and are indicated as downward arrows.  Four galaxies do not
  have sensitive constraints for $\mathrm{\langle
    \mathit{N}(Fe\,II)/\mathit{N}(Mg\,II)\rangle}$, either due to
  heavily saturated Mg\,II absorbers or absence of a strong limit on
  the Fe\,II absorption.  These data points are greyed out for
  clarity.  Following Figure 3, the dotted horizontal line indicates
  the solar (Fe/Mg) abundance pattern, for visual comparisons.  Two
  interesting features are evident: (1) passive galaxies exhibit on
  average higher $\mathrm{\langle
    \mathit{N}(Fe\,II)/\mathit{N}(Mg\,II)\rangle}$ than star-forming
  galaxies; and (2) $\mathrm{\langle
    \mathit{N}(Fe\,II)/\mathit{N}(Mg\,II)\rangle}$ appears to show a
  mild declining trend toward larger $d$ for both passive and
  star-forming subsamples.}
\label{Figure 4}
\end{figure}

\section[]{The radial profile of $\bf \it{N}({\rm Fe\,II})/\it{N}({\rm Mg\,II})$ in galaxy halos}

Figure 3 suggests that $N({\rm Fe\,II})/N({\rm Mg\,II})$ in galaxy
halos depends on both the projected distance and star formation
history of the absorbing galaxy.  To better quantify how the observed
$N({\rm Fe\,II})/N({\rm Mg\,II)}$ ratio depends on galaxy properties,
we compute a $N$(Mg\,II)-weighted mean Fe\,II to Mg\,II column density
ratio for each absorption system according to the following equation,
\begin{align}
%\mathrm{log\,\left \langle \frac{\mathit{N}(Fe\,II)}{\mathit{N}(Mg\,II)} \right \rangle_{tot}} 
\mathrm{log\,\left \langle \frac{\mathit{N}(Fe\,II)}{\mathit{N}(Mg\,II)} \right \rangle} 
&= \mathrm{log\, \sum \limits_{\mathit{i}} \mathit{w_i}\dfrac{\mathit{N_i}(Fe\,II)}{\mathit{N_i}(Mg\,II)}} \nonumber\\
&= \mathrm{log\,\mathit{N}_{tot}(Fe\,II)}-\mathrm{log\,\mathit{N}_{tot}(Mg\,II)} ,
\end{align}
where $w_i= N_i\mathrm{(Mg\,II)}/N_\mathrm{tot}\mathrm{(Mg\,II)}$ for
component $i$, and log\,$N_\mathrm{tot}$(Fe\,II) and
log\,$N_\mathrm{tot}$(Mg\,II) are the total column densities of Fe\,II
and Mg\,II, respectively, summed over all individual absorbing
components in a given system. The mean $N({\rm Fe\,II})/N({\rm
  Mg\,II)}$ ratio for each absorption system is presented in Column 11
of Table 1.  The quoted uncertainty in $\mathrm{log\,\langle
  \mathit{N}(Fe\,II)/\mathit{N}(Mg\,II) \rangle}$ for each system is
the dispersion from the weighted mean, which represents the scatter in
Fe\,II to Mg\,II column density ratio among individual components of
the corresponding absorber.

Figure 4 displays $\mathrm{\langle
  \mathit{N}(Fe\,II)/\mathit{N}(Mg\,II)\rangle}$ versus $d$ for
passive (red circles) and star-forming (blue stars) galaxies.  Three
galaxies exhibit heavily saturated Mg\,II absorbers: (1) the
passive galaxy at $z=0.5437$ and at $d=74$ kpc from
SDSSJ\,1146$+$0207; (2) the passive galaxy at $z=0.465$ and at $d=166$ kpc from SDSSJ\,1025$+$0349; and (3) the star-forming galaxy at $z=0.892$ and at $d=23$ kpc from 3C336.  No meaningful constraints for
$\mathrm{\langle \mathit{N}(Fe\,II)/\mathit{N}(Mg\,II)\rangle}$ can be
obtained for these absorbers.  In addition, the star-forming galaxy at
$z=0.3404$ and at $d=85$ kpc from PKS\,1424$-$1150 exhibit only a weak
Mg\,II absorber and Fe\,II absorption is not detected.  The available
absorption spectra do not place a sensitive constraint for
$\mathrm{\langle \mathit{N}(Fe\,II)/\mathit{N}(Mg\,II)\rangle}$. All
four galaxies are greyed out in Figure 4 for clarity.  Considering
only galaxies for which measurements of (or strong constraints for)
$\mathrm{\langle \mathit{N}(Fe\,II)/\mathit{N}(Mg\,II)\rangle}$ are
available, it is clear that passive galaxies exhibit on average higher
$\mathrm{\langle \mathit{N}(Fe\,II)/\mathit{N}(Mg\,II)\rangle}$ than
star-forming galaxies.  In addition, $\mathrm{\langle
  \mathit{N}(Fe\,II)/\mathit{N}(Mg\,II)\rangle}$ appears to show a
mild declining trend toward larger $d$ for both passive and
star-forming subsamples.

To further examine the dependence of $N({\rm Fe\,II})/N({\rm Mg\,II)}$
ratio on galaxy projected distance $d$, we divide each of the passive
and star-forming subsamples into low- and high-$d$ subsamples.  The
adopted bin size in $d$ is determined so that there are roughly equal
number of galaxies (five to seven galaxies) in each subsample.  We
then compute the median value $\mathrm{log\,\langle
  \mathit{N}(Fe\,II)/\mathit{N}(Mg\,II)}\rangle_{\rm med}$ in each bin
as well as the dispersion of each subsample around the median value.
Saturated absorption systems are excluded from this exercise, because
no constraints on $N({\rm Fe\,II})/N({\rm Mg\,II)}$ can be derived
either due to heavily saturated Mg\,II absorption or insufficient
limits on $N({\rm Fe\,II})/N({\rm Mg\,II})$ (see Column 12 of Table 1). In addition, for passive galaxies at $d\apg 100$ kpc two of the seven galaxies have only a relatively strong upper limit at $\mathrm{log\,\langle
  \mathit{N}(Fe\,II)/\mathit{N}(Mg\,II)}\rangle\approx-0.3$.  For
this subsample, we infer a 85\% upper limit for the underlying
distribution of $\mathrm{log\,\langle
  \mathit{N}(Fe\,II)/\mathit{N}(Mg\,II)}\rangle_{\rm med}<-0.3$.  The
results are presented in Figure 5 for passive galaxies (red circles)
and star-forming galaxies (blue stars).

Figure 5 shows that $\mathrm{log\,\langle
  \mathit{N}(Fe\,II)/\mathit{N}(Mg\,II)}\rangle_{\rm med}$ declines
with increasing $d$, from $\mathrm{log\,\langle
  \mathit{N}(Fe\,II)/\mathit{N}(Mg\,II)}\rangle_{\rm med} = -0.06\pm
0.15$ at $d<60$ kpc to $\mathrm{log\,\langle
  \mathit{N}(Fe\,II)/\mathit{N}(Mg\,II)}\rangle_{\rm med}< -0.3$ at
$d>100$ kpc.  For star-forming galaxies, a similar declining trend is
found with increasing $d$, from $\mathrm{log\,\langle
  \mathit{N}(Fe\,II)/\mathit{N}(Mg\,II)}\rangle_{\rm med} = -0.25\pm
0.21$ at $d<44$ kpc to $\mathrm{log\,\langle
  \mathit{N}(Fe\,II)/\mathit{N}(Mg\,II)}\rangle_{\rm med} = -0.91\pm
0.44$ at $d>70$ kpc.  Furthermore, while $\mathrm{log\,\langle
  \mathit{N}(Fe\,II)/\mathit{N}(Mg\,II)}\rangle_{\rm med}$ is about
0.2 dex higher in the inner halos ($d\apl 60$ kpc) of passive galaxies
than those around star-forming ones, both galaxy populations display a
comparable $\mathrm{log\,\langle
  \mathit{N}(Fe\,II)/\mathit{N}(Mg\,II)}\rangle_{\rm med}$ in the
outer halos.

\begin{figure} 
\hspace{-0.5em}
\includegraphics[width=87mm]{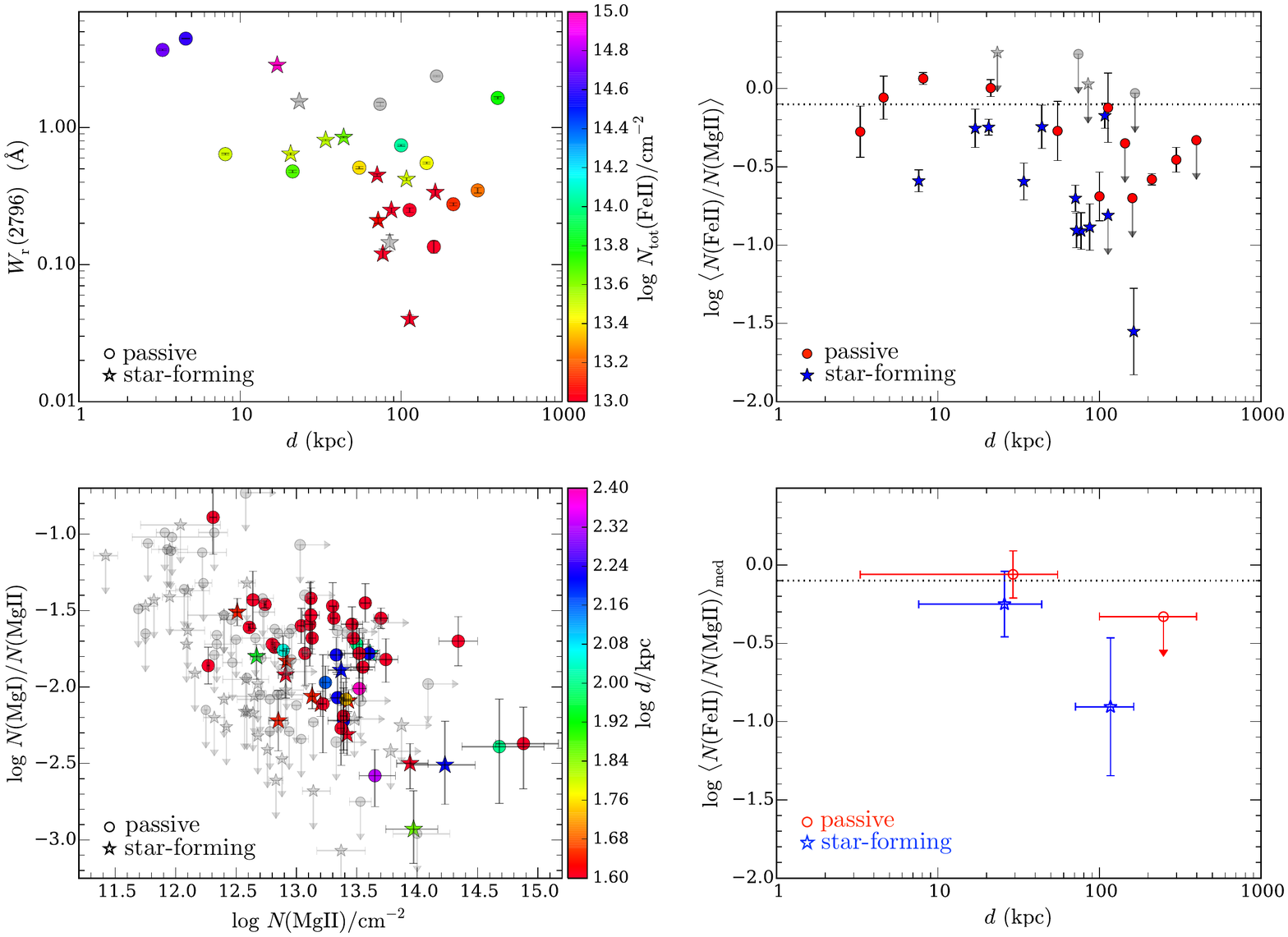}
\vspace{-1.5em}
\caption{Dependence of median column density ratio,
  $\mathrm{log\,\langle
    \mathit{N}(Fe\,II)/\mathit{N}(Mg\,II)}\rangle_{\rm med}$, on
  galaxy projected distance $d$ for passive galaxies (red circles) and
  star-forming galaxies (blue stars).  The horizontal error bars
  represent the full range of projected distances in each bin, whereas
  the vertical error bars show the dispersion from
  $\mathrm{log\,\langle
    \mathit{N}(Fe\,II)/\mathit{N}(Mg\,II)}\rangle_\mathrm{med}$ in
  each bin.  Two of the seven passive galaxies at $d\apg 100$ kpc have
  only a relatively strong upper limit at $\mathrm{log\,\langle
    \mathit{N}(Fe\,II)/\mathit{N}(Mg\,II)}\rangle \approx-0.3$ (Figure
  4).  For this subsample, we infer a 85\% upper limit for the
  underlying distribution of $\mathrm{log\,\langle
    \mathit{N}(Fe\,II)/\mathit{N}(Mg\,II)}\rangle_{\rm med}<-0.3$,
  which is shown as the downward arrow.  Like Figure 3, the dotted
  horizontal line indicates the solar (Fe/Mg) abundance pattern.}
\label{Figure 5}
\end{figure}

\section[]{Discussion}

A primary goal of our study is to determine the extent
SNe~Ia-dominated feedback in gas around galaxies.  The analysis
presented in \S\S\ 3 \& 4 indicate that passive galaxies display on
average a higher $\mathrm{\mathit{N}(Fe\,II)/\mathit{N}(Mg\,II)}$ in
their halos than star-forming galaxies.  In addition, there also
appears to be a modest decline in the relative ionic ratio with
increasing projected distance.  To infer the underlying elemental
abundance ratio between Fe and Mg from the observed relative
abundances of Fe$^+$ and Mg$^+$, it is necessary to first address the
differential ionization fraction between the two ions.  Furthermore,
it is also necessary to quantify possible systematic biases due to
differential dust depletion.  Here we discuss both of these effects
and the implications of our findings.

\subsection{Differential ionization fraction}

To determine the ionization state of the gas, measurements of $N({\rm
  H\,I})$ and relative abundance ratios between multiple ions are
often necessary.  While the observed relative ionic ratios constrain
the ionization parameters, knowledge of $N({\rm H\,I})$ determines
whether the gas is optically-thin or optically-thick to the background
radiation field.  However, $N({\rm H\,I})$ is not known for all but
one galaxy in our sample.  As illustrated in Zahedy \etal\ (2016),
useful empirical constraints on the relative Fe/$\alpha$ abundances
can be obtained in the absence of $N({\rm H\,I})$.  This is achieved
by performing a series of photoionization calculations that explore a
wide range of ionization conditions.  Then, constraints for the
ionization state of the gas are obtained based on comparisons of the
predicted and observed column density ratios between $\mathrm{Mg^0}$
and $\mathrm{Mg^+}$ ions.

In Figure 6 we present the observed Mg\,I to Mg\,II column density
ratio, $N({\rm Mg\,I})/N({\rm Mg\,II})$, versus Mg\,II column density,
$N({\rm Mg\,II})$, for individual absorbing components associated with
passive (circle symbols) and star-forming (star symbols) galaxies.
The data points are color-coded by the projected distances of the
absorbing galaxies.  Grey data points represent 2-$\sigma$ upper limit
of $N({\rm Mg\,I})/N({\rm Mg\,II)}$, due to either non-detection of
Mg\,I or saturated Mg\,II absorption.  Figure 6 shows that the
majority ($>90\%$) of absorbing components with log\,$N
\mathrm{(Mg\,II)} \apl 13.8$ occur in a range of log\,$N({\rm
  Mg\,I})/N({\rm Mg\,II})$ values of $-2.2 \apl \mathrm{log}\,N({\rm
  Mg\,I})/N({\rm Mg\,II}) \apl -1.4$.  On the other hand, for high
column density absorbing components, log\,$N \mathrm{(Mg\,II)}\apg14$,
the ratios are lower with a typical $\mathrm{log}\,N({\rm
  Mg\,I})/N({\rm Mg\,II}) \sim -2.5$, albeit with larger
uncertainties.  The range of $N({\rm Mg\,I})/N({\rm Mg\,II})$ in
Figure 6 is consistent with what has been found for randomly selected
Mg\,II absorbers by Churchill \etal\ (2003).

In Zahedy \etal\ (2016), we performed a series of photoionization
calculations for a photoionized gas of temperature $T=10^4$ K and a
range of gas densities, metallicities, and $N({\rm H\,I})$.  A
plane-parallel geometry was assumed for the gas, which was illuminated
on both sides with the updated Haardt \& Madau (2001) ionizing
background radiation field (HM05 in \textsc{Cloudy} v.13.03; Ferland
\etal\ 2013) at $z = 0.5$.  For each model, the expected relative
abundance ratio between $\mathrm{Mg^0}$ and $\mathrm{Mg^+}$ ions and
the ionization fraction ratio between $\mathrm{Fe^+}$ and
$\mathrm{Mg^+}$ were calculated for a gas that follows the solar
abundance pattern.  For absorption components with log\,$N
\mathrm{(Mg\,II)} \apl 13.8$ in the present study, comparing the
observed $N({\rm Mg\,I})/N({\rm Mg\,II})$ and predictions from Zahedy
\etal\ (2016) leads to constraints on the gas density of
$3\times10^{-2}\, \cmjjj \apl n_\mathrm{H} \apl 2\times 10^{-3}\,
\cmjjj$.  Over the range of allowed gas densities, we find that the
ionization fraction of Fe$^+$ ($f_{{\rm Fe}^+}$) remains roughly equal
to that of Mg$^+$ ($f_{{\rm Mg}^+}$) in the optically-thick regime and
lower than Mg$^+$ in optically-thin gas.  As a result, $f_{{\rm
    Fe}^+}/f_{{\rm Mg}^+}<1$ and $N({\rm Fe\,II})/N({\rm Mg\,II})$,
which is equal to $(f_{{\rm Fe}^+}/f_{{\rm Mg}^+})\times\,N({\rm
  Fe})/N({\rm Mg})$, marks a lower limit to the underlying $N({\rm
  Fe})/N({\rm Mg})$.

For strong Mg\,II components of log\,$N \mathrm{(Mg\,II)} \apg 14$,
however, the observed low $N({\rm Mg\,I})/N({\rm Mg\,II})$ ratios
require a low density gas of $n_\mathrm{H} \approx 10^{-3}\, \cmjjj$
for the same HM05 ionizing radiation intensity.  The inferred low gas
density together with a large $N \mathrm{(Mg\,II)}$ implies a cloud
size of $>10$ kpc if the gas has a solar metallicity or $>100$ kpc if
the gas metallicity is 0.1 solar.  The unphysically large cloud sizes
inferred for the strongest Mg\,II absorbing components raise questions
for the accuracy of the photoionization models.

Because the observed ionic ratios in photoionization models is
dictated by the ionization parameter, which is the number of ionizing
photons per atom, a natural explanation for the inferred low gas
density is that the ionizing radiation intensity has been
underestimated.  To increase the ionizing radiation intensity from the
standard HM05, we experiment with adding a local ionizing radiation
field due to the absorbing galaxy.  We first generate a synthetic
galaxy spectrum using Starburst99 (Leitherer \etal\ 1999) and assuming
a star formation rate of 1 $M_\odot\,\mathrm{yr}^{-1}$, an age of
$10^7\,\mathrm{yr}$, and an ionizing photon escape fraction of 2\%.
Then, we normalize the flux to match an $L_*$ galaxy at $z=0.5$.  For a
gas cloud located at 15 kpc from the galaxy, we find that the ionizing
radiation from the galaxy provides a 20-fold increase to the number
density of hydrogen-ionizing photons.  With an increased ionizing
radiation intensity, the inferred underlying gas density increases
accordingly.  We find that with this revised radiation field the
observed low $N({\rm Mg\,I})/N({\rm Mg\,II})$ for strong Mg\,II
components of log\,$N \mathrm{(Mg\,II)} \apg 14$ can be reproduced for
a gas density of $n_\mathrm{H}\sim 0.01\, \cmjjj$.  At this gas
density, we find that log\,$f_{{\rm Fe}^+}/f_{{\rm Mg}^+}\approx -0.4$
and confirm that the observed $N({\rm Fe\,II})/N({\rm Mg\,II})$
represents a lower limit to the underlying $N({\rm Fe})/N({\rm Mg})$.

An alternative explanation for the observed low $N({\rm Mg\,I})/N({\rm
  Mg\,II})$ in strong Mg\,II absorbing components is the presence of
additional heating sources that may increase the ionization of the gas
in a warmer temperature regime.  To explore this alternative scenario,
we repeat the photoionization calculations for a higher temperature of
$T=3\times10^4$ K, assuming the original HM05 radiation field.  The
adopted higher temperature is motivated by the observed Doppler
parameters of individual Mg\,II, Mg\,I and Fe\,II absorption
components.  A median value of $b\sim5$ \kms\ is found for each of the three
transitions, placing an upper limit on the allowed gas temperature at
$T \apl 3\times10^4$ K.

The resulting model predictions show that the observed log\,$N({\rm
  Mg\,I})/N({\rm Mg\,II})\sim-2.5$ in most log\,$N \mathrm{(Mg\,II)}
\apg 14$ absorbing components can also be reproduced for a warm
($T=3\times10^4$ K) optically-thick gas of $\log\,N({\rm H\,I})\apg
19$ and 0.1 solar metallicity under a standard HM05 radiation field
over a wide range of gas densities.  For gas with lower-$N({\rm
  H\,I})$, a still higher gas temperature is required, which then
becomes incompatible with the observed $b$ value.  Imposing a maximum
cloud size of $l\sim10$ kpc based on observations of Galactic
high-velocity clouds (e.g., Putman \etal\ 2012) constrains the gas
density to be $n_\mathrm{H}\apg 4\times 10^{-3}\, \cmjjj$ for a gas
with 0.1 solar metallicity.  Over the range of allowed $n_\mathrm{H}$,
log\,$f_{{\rm Fe}^+}/f_{{\rm Mg}^+} \approx -0.3$, indicating that the
observed $N({\rm Fe\,II})/N({\rm Mg\,II})$ marks a lower limit to the
underlying $N({\rm Fe})/N({\rm Mg})$.

In summary, the exercise described in this section demonstrates that
\emph{for both weak and strong Mg\,II absorption systems identified in
  this study, the observed}
$\mathrm{\mathit{N}(Fe\,II)/\mathit{N}(Mg\,II)}$ \emph {ratio
  represents a lower limit to the underlying elemental} (Fe/Mg) \emph{
  ratio of the gas for both optically-thin and thick gases}.  Comparing the observed $\mathrm{\mathit{N}(Fe\,II)/\mathit{N}(Mg\,II)}$
to the solar (Fe/Mg) abundance ratio from Asplund \etal\ (2009; dotted
line in Figures $3-5$), our analysis shows that $>60$\% of passive
galaxies exhibit an enhanced Fe/$\alpha$ elemental abundance ratio at
$d\apl 60$ kpc that exceeds what is observed in the solar
neighborhood.  At the same time, the majority of passive galaxies at
$d\apg 100$ kpc and all star-forming galaxies at $d\apl 150$ kpc
exhibit chemical compositions consistent with $\alpha$-element
enhancement.

\begin{figure} 
\hspace{-0.5em}
\includegraphics[width=86mm]{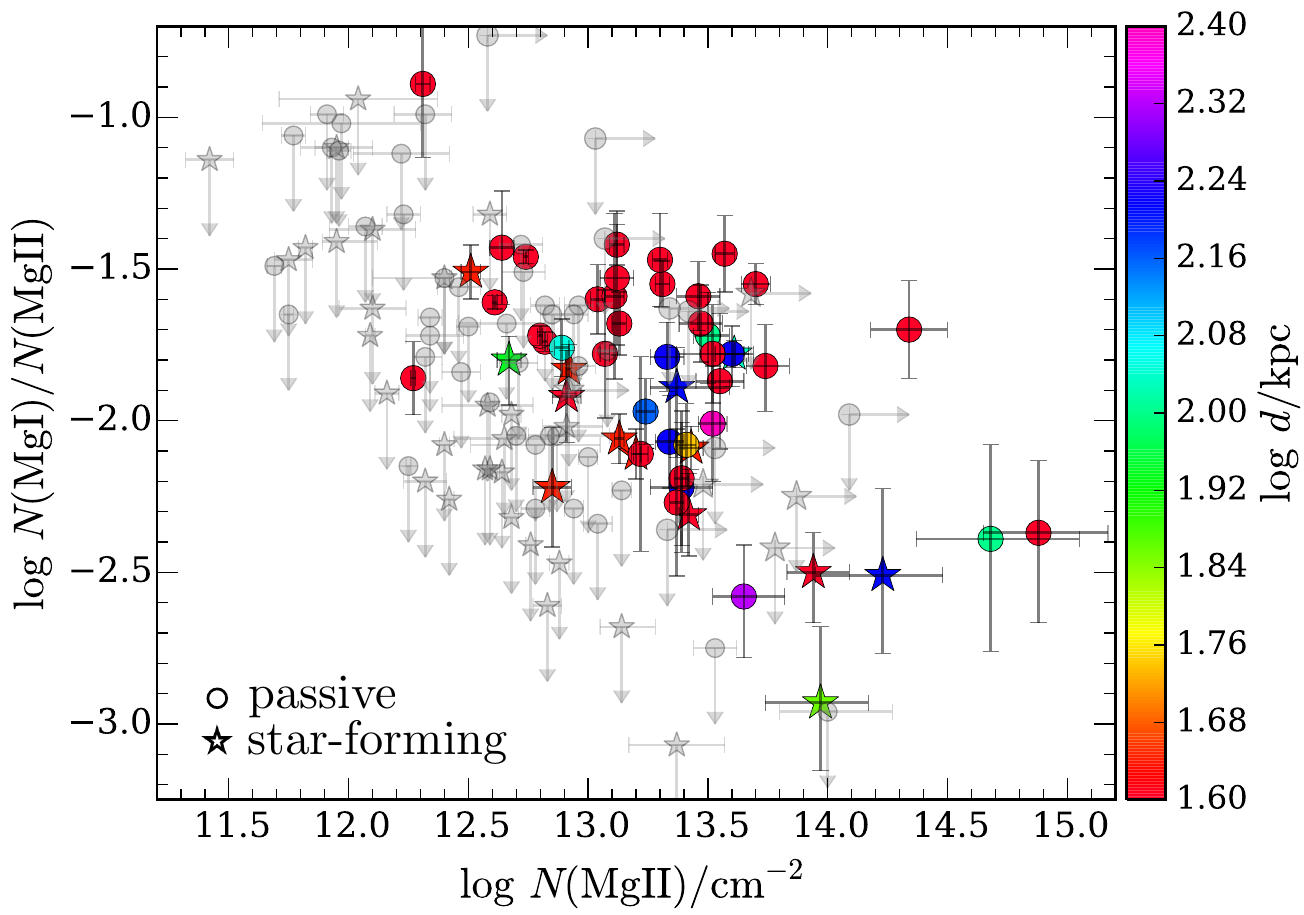}
\vspace{-1.5em}
\caption{Observed Mg\,I to Mg\,II column density ratio, $N({\rm
    Mg\,I})/N({\rm Mg\,II})$, versus Mg\,II column density, $N({\rm
    Mg\,II})$, for individual components identified in our Voigt
  profile analysis (\S\ 3).  Absorption components associated with
  passive galaxies are plotted in circle symbols, whereas those
  associated with star-forming galaxies are plotted in star
  symbols. The error bars associated with each data point show the
  measurement uncertainties.  The color of each data point represents
  the projected distance $d$ of the absorbing galaxy to the background
  QSO sightline.  When a Mg\,II absorbing component has no
  corresponding Mg\,I absorption detected, it is shown as a downward
  arrow, with the gray data point indicating the 2-$\sigma$ upper
  limit of $N({\rm Mg\,I})/N({\rm Mg\,II)}$.}
  \label{Figure 5}
\end{figure}

\subsection[]{Differential dust depletion}

It is known from observations of the Milky Way ISM that Fe is more
readily incorporated into dust grains than Mg, showing in excess of
0.5 dex more depletion than Mg in cool ($T\sim 500$ K) ISM gas (e.g,
Savage \& Sembach 1996).  Including such differential dust depletion,
the inferred $N({\rm Fe})/N({\rm Mg})$ after accounting for possible
differential ionization fraction correction may still represent a
lower limit to the underlying elemental abundance ratio.  At the same
time, dust grains are expected to be easily destroyed in warmer
environment (e.g., Draine \& Salpeter 1979).  Indeed, the observed
differential depletion between Fe and Mg reduces to $\approx 0.35$ dex
in warm ($T\approx 6000$ K) ISM gas (e.g, Savage \& Sembach 1996).
For Mg\,II absorbing gas of $T\sim 10^4$ K, it is expected that dust
destruction is still more effective (e.g., McKee \etal\ 1987) and that
differential dust depletion is at its minimum.  X-ray observations of
the hot ISM gas in 19 local early-type galaxies from Humphrey \& Buote
(2006) have yielded a median Fe to Mg ratio of
log\,$\mathrm{(Fe/Mg)}=-0.01\pm0.18$ at $d \apl 10-60$ kpc from these
galaxies.  This median value is consistent with what we found for
singly ionized Fe and Mg at $d<60$ kpc from $z\sim0.5$ passive
galaxies, supporting the expectation that dust depletion is not
significant in the cool ($T\sim 10^4$ K) halo clouds revealed by
Mg\,II absorption transitions.

However, it is possible that the observed low
$\mathrm{\mathit{N}(Fe\,II)/\mathit{N}(Mg\,II)}$ ratio around
star-forming galaxies is due to a larger amount of differential dust
depletion around these galaxies, particularly for the Mg\,II absorbers
detected at $d<50$ kpc.  These absorbers are all relatively strong
with $W_r\mathrm{(2796)}\apg 0.6\,\mathrm{\AA}$, and it has been shown
that $\approx 30-40$\% of these strong Mg\,II absorbers contain
neutral gas of $N{\rm (H\,I})\apg 2\times 10^{20} \cmjj$ (e.g., Rao
\etal\ 2006).  Therefore, a large fraction of these absorbers are
likely to arise in DLA gas.  For DLAs at $z\sim0.5-3$, Fe is found to
be more depleted than Mg by $\approx 0.2$ dex (e.g., Vladilo
\etal\ 2011; De Cia \etal\ 2016).  If dust is present in these strong
Mg\,II absorbers around $z\sim 0.5$ star-forming galaxies, then
adopting the mean differential dust depletion between Fe and Mg
observed for DLAs would imply a median underlying (Fe/Mg) abundance
ratio of log\,$\mathrm{(Fe/Mg)}=-0.05$ at $d<50$ kpc, comparable to
both the solar value and what is observed at $d<60$ kpc from passive
galaxies.  For weaker Mg\,II absorbers at $d>50$ kpc, the dust content
is expected to be significantly less and little correction for
differential dust depletion is expected.

The differential dust depletion between Fe and Mg can be inferred
directly for the absorption systems around two star-forming galaxies
in the $d<50$ kpc subsample, providing a comparison to the underlying
(Fe/Mg) ratio implied from the subsample median.  For the star-forming
galaxy at $z = 0.101$ and $d = 8$ kpc from PKS\,0439$-$433, the
roughly solar metallicity known for the DLA gas associated with this
galaxy (Chen \etal\ 2005; Som \etal\ 2015) allows us to estimate the
dust depletion factors for Fe and Mg based on a known correlation
between dust depletion and gas-phase metallicity.  The differential
dust depletion between Fe and Mg is expected to vary between $+0.5$
and $+0.8$ dex for a solar-metallicity gas, based on observations of
Galactic absorbers (e.g., De Cia \etal\ 2016).  If dust is present in
this DLA, this range of differential dust depletion would imply an
underlying (Fe/Mg) abundance ratio of log\,$\mathrm{(Fe/Mg)}=-0.1$ to
$+0.2$ for the gas.  Similarly for the ultra-strong Mg\,II absorber
associated with the $z=1.24$ galaxy at $d = 17$ kpc from
SDSSJ\,1430$+$0149, although the metallicity is not known, the
relative abundance ratio of Cr to Zn is estimated to be
[Cr/Zn]$=-0.51\pm0.06$ (Zych \etal\ 2009), indicating a modest level
of dust depletion comparable to what is seen in the Galactic Halo
(e.g., Savage \& Sembach 1996). The expected differential depletion
between Fe and Mg in such an environment is $\approx+0.2$ dex, which
would imply an underlying (Fe/Mg) ratio of
log\,$\mathrm{(Fe/Mg)}=-0.05$ for this Mg\,II absorber.  In both of
these cases, applying the estimated dust depletion correction
individually for each galaxy results in an implied underlying (Fe/Mg)
abundance ratio that is comparable to solar value, consistent with
what is found by applying the mean dust depletion correction for DLAs
on the median value for the inner-$d$ bin.  This exercise lends strong
support for the finding that solar-level (Fe/Mg) gas may not be
uncommon at $d<50$ kpc from intermediate-redshift, star-forming
galaxies.
 
\subsection[]{Implications on the origin of chemically enriched gas in galaxy halos}

As previously mentioned in \S\ 1, iron is produced in both core
collapse and Type Ia SNe, whereas magnesium is produced primarily in
massive stars and core-collapse SNe. Specifically, a Type Ia supernova
is expected to produce $\sim0.7\,M_\odot$ of Fe while releasing no
more than $0.02\,M_\odot$ of magnesium at the same time (e.g., Iwamoto
\etal\ 1999).  Different types of SNe originate in different
progenitor stars of different stellar ages.  The elemental (Fe/Mg)
ratio therefore provides a measure of the relative contributions from
different massive stars to the chemical enrichment history of a galaxy
(e.g., de Plaa \etal\ 2007; Zahedy \etal\ 2016).  Consequently, the
inferred lower limit of the underlying (Fe/Mg) elemental abundance
ratio from the observed $N({\rm Mg\,I})/N({\rm Mg\,II)}$ provides a
useful ``clock'' for timing the age of the stellar population.  In
addition, the spatial profiles of $N({\rm Mg\,I})/N({\rm Mg\,II)}$
offers important constraints for the extent of SNe Ia-dominated
chemical enrichment in galactic halos.

For quiescent galaxies in this study, the observed
$\mathrm{\mathit{N}(Fe\,II)/\mathit{N}(Mg\,II)}$ at $d \apl 60$ kpc is
high with a median and dispersion of $\mathrm{log\,\langle
  \mathit{N}(Fe\,II)/\mathit{N}(Mg\,II)}\rangle_\mathrm{med}=-0.06\pm0.15$.
The large column density ratio implies a lower limit to the underlying
(Fe/Mg) ratio of the gas at $[{\rm Fe}/{\rm Mg}]\equiv\log\,({\rm
  Fe}/{\rm Mg})-\log\,({\rm Fe}/{\rm Mg})_\odot\apg 0$. The implied
fractional contribution from SNe Ia to the chemical enrichment of the
gas is $f_\mathrm{Ia}\apg 17\%$ based on the expected
nucleosynthetic yields for Type Ia and core-collapse SNe from Iwamoto
\etal\ (1999). This minimum value is comparable to what has been
estimated for solar-abundance gas in the Milky Way (e.g., Tsujimoto
\etal\ 1995). With such a significant contribution from SNe Ia, it can
also be expected that cool gas at $d\apl60$ kpc from passive galaxies
has been enriched to a relatively high metallicity, reflecting the
role of multiple generations of massive stars and SNe Ia in its
chemical enrichment history. Indeed, this expectation is at least
consistent with what has been found in the hot ISM of local elliptical
galaxies, where near-solar mean metallicities are commonly observed at
a similar range of $d$ (e.g., Humphrey \& Buote 2006; Loewenstein \&
Davis 2010, 2012). Furthermore, we note that Fe/$\alpha$ radial
profile measurements are available from X-ray observations of the hot
ISM in several nearby massive quiescent galaxies (e.g, Nagino \&
Matsushita 2010; Loewenstein \& Davis 2010, 2012), where it has been
found that the Fe/$\alpha$ radial profile in these galaxies is
consistent within measurement errors with being flat at
$[\mathrm{Fe}/\alpha]\sim0$ level at $d \apl 40$ kpc, comparable to
what can be inferred from our observations at $d \apl 60$ kpc from
$z\sim 0.5$ passive galaxies.

Our analysis also provides a quantitative constraint on the
Fe/$\alpha$ ratio at $d \apg 100$ kpc from passive galaxies, where the
gas is typically too diffuse to be detected in emission even in the
local universe.  
%While at $d\apl50$ kpc both cool gas revealed in
%absorption and hot ISM gas detected in X-ray emission have
%$\mathrm{\mathit{N}(Fe\,II)/\mathit{N}(Mg\,II)}$ values that are
%consistent with Fe-rich (Fe/Mg) ratios, it is interesting that at
We find that at $d\apg100$ kpc the absorbing gas generally shows lower
$\mathrm{\mathit{N}(Fe\,II)/\mathit{N}(Mg\,II)}$ ratios than typically seen at $d\apl60$ kpc.  Specifically
for passive galaxies, the
$\mathrm{\mathit{N}(Fe\,II)/\mathit{N}(Mg\,II)}$ ratios at $d\apg100$
kpc can be characterized by a 85\% upper limit to the underlying
distribution of $\mathrm{log\,\langle
  \mathit{N}(Fe\,II)/\mathit{N}(Mg\,II)}\rangle_{\rm med}<-0.3$.  In
at least three out of seven cases ($\approx40\%$), the absorbing gas
displays $\mathrm{log\,\langle
  \mathit{N}(Fe\,II)/\mathit{N}(Mg\,II)}\rangle_{\rm
  med}\apl-0.6$. Allowing a modest differential ionization fraction
between $\mathrm{Fe}^+$ and $\mathrm{Mg}^+$ of up to log $f_{{\rm
    Fe}^+}/f_{{\rm Mg}^+}\approx-0.3$ dex (assuming an optically thick
gas, see \S\ 5.2), the implied underlying (Fe/Mg) relative abundance
ratio is $[\mathrm{Fe/Mg}]\apl-0.2$ for a
log\,$\mathrm{\mathit{N}(Fe\,II)/\mathit{N}(Mg\,II)=-0.6}$ gas.  This
range of $[\mathrm{Fe/Mg}]$ ratios is comparable to that of
high-redshift DLAs, where the mean $[\mathrm{Fe}/\alpha]$ has been
found to be $-0.26\pm0.12$ at $z\apg 3$ (e.g., Rafelski
\etal\ 2012). For $[\mathrm{Fe/Mg}]\apl-0.2$, the expected maximum
fractional contribution from SNe Ia to the chemical enrichment is
$f_\mathrm{Ia}\apl 5 \%$, based on supernova nucleosynthetic
yields from Iwamoto \etal\ (1999).  Our analysis therefore indicates
that SNe~Ia-driven chemical enrichment is relatively localized in
inner halos at $d\apl 60$ kpc and that the chemical enrichment of the
cool gas at $d\apg100$ kpc from $z=0.5$ passive galaxies is consistent
with an early enrichment driven by core-collapse SNe.

For star-forming galaxies, interpretations of our observations are
more uncertain.  As previously noted in \S\ 5.2, if differential dust
depletion is important, then the absorbing gas at $d\apl 50$ kpc from
these star-forming galaxies may also have a near solar-level (Fe/Mg)
ratio.  Such a high level Fe/Mg ratio may not be surprising for a
mature disk galaxy at low- to intermediate-redshifts.  However, it is
interesting to find that a strong Mg\,II absorber of $W_r(2796)=2.86$
at $d=17$ kpc from a star-forming galaxy at $z=1.24$ could potentially
have a solar Fe/Mg ratio after differential dust depletion
corrections.  The implication is that the galaxy contains a relatively
evolved stellar population, while maintaining an active level of star
formation when the universe was $\approx 5$ Gyr old.  Follow-up
studies of the stellar populations and star formation history in this
galaxy are necessary for a better understanding of the observed high
Fe/Mg ratio in the absorber.

At $d>70$ kpc from star-forming galaxies, the observed Fe\,II to Mg\,II column density ratios are generally low with a median and dispersion of $\mathrm{log\,\langle \mathit{N}(Fe\,II)/\mathit{N}(Mg\,II)}\rangle_{\rm
  med}=-0.91\pm0.44$.  For a median $N({\rm Mg\,I})/N({\rm Mg\,II})$
ratio of log\,$N({\rm Mg\,I})/N({\rm Mg\,II})=-1.9$, the expected
differential ionization fraction between $\mathrm{Fe}^+$ and
$\mathrm{Mg}^+$ is log $f_{{\rm Fe}^+}/f_{{\rm Mg}^+}\approx-0.6$ dex
for an optically-thin gas (and negligible for an optically-thick gas).
Applying this ionization correction leads to an implied underlying
(Fe/Mg) relative abundance ratio of $[\mathrm{Fe/Mg}]\apl-0.2$ at
$d=70-160$ kpc. Similar to what is seen at $d>100$ kpc from passive
galaxies, cool gas at $d>70$ kpc from star-forming galaxies also shows
an $\alpha$-element enhanced abundance pattern driven by core-collapse
SNe.

In summary, our analysis suggests that Fe/$\alpha$ in galactic halos
declines with increasing projected distance from both passive and
star-forming galaxies.  At $d\apl 60$ kpc, a significant contribution
from SNe~Ia ($f_\mathrm{Ia}\apg 15$\%) is necessary to explain the
observed Fe/Mg ratios, whereas at $d\apg 70$ kpc, contributions from
SNe~Ia are limited to $f_\mathrm{Ia}\apl 5$\% in both star-forming and
quiescent halos.  Together, our analysis shows that SNe~Ia-driven
chemical enrichment is relatively localized in inner halos at $d\apl
60$ kpc.  Alternatively, inflowing gas from the IGM could also
``dilute'' an Fe-rich gas and produce a declining Fe/$\alpha$ radial
profile in the halo, because accreted IGM gas is expected to show an
$\alpha$-enhanced abundance pattern which reflects the early
enrichment history (e.g., Rauch \etal\ 1997).  However, such dilution
effect is likely minimal if the metallicity of an $\alpha$-enhanced
inflowing gas is significantly lower than the Fe-rich gas.

It is clear that there is a significant scatter in the empirical
measurements of $N({\rm Fe\,II})/N({\rm Mg\,II})$, while the inferred
$f_\mathrm{Ia}$ depends sensitively on the input Fe/$\alpha$ elemental
abundance ratio.  We anticipate that future observations combining
metallicity and relative abundance measurements in stars/ISM and in
halo gas for a large sample of galaxies will provide the precision
necessary to distinguish between different scenarios.

\section*{Acknowledgments}

We thank the anonymous referee for thoughtful comments which helped us improve the paper. The authors thank Sean Johnson and John Mulchaey for useful discussions and suggestions, and Yun-Hsin Huang, and Sean Johnson for obtaining some of the MIKE spectra analyzed here.  This work is based
on data gathered with the 6.5 m Magellan Telescopes located at Las
Campanas Observatory, the ESO telescopes at the La Silla Paranal
Observatory, and the NASA/ESA Hubble Space Telescope operated by the
Space Telescope Science Institute and the Association of Universities
for Research in Astronomy, Inc., under NASA contract NAS 5-26555.
Additional data were obtained at the W.M. Keck Observatory, which is
operated as a scientific partnership among the California Institute of
Technology, the University of California and the National Aeronautics
and Space Administration.  The Observatory was made possible by the
generous financial support of the W.M. Keck Foundation.

\bsp

\label{lastpage}

\end{document}